\documentclass[12pt] {article}
\pdfoutput=1
\usepackage[hyphens,spaces,obeyspaces]{url}
\usepackage{setspace}

\usepackage{epsfig}
\usepackage{amssymb}
\usepackage{amsmath}
\usepackage{multirow}
\usepackage{setspace}
\usepackage{datetime}
\usepackage{amsmath}
\usepackage{algorithm}
\usepackage{algpseudocode}
\usepackage{graphicx}
\usepackage{lipsum}
\usepackage{acronym}
\usepackage{array}
\usepackage{graphicx}
\usepackage{cite}
\usepackage{graphicx}
\usepackage{epstopdf}
\usepackage{caption}
\usepackage{balance}
\usepackage{fancybox}
\usepackage{colortbl}
\usepackage{array}
\usepackage{mystyle}
\usepackage{multirow}
\usepackage[obeyspaces,hyphens,spaces]{url}
\usepackage{threeparttable}
\usepackage[english]{babel}
\usepackage{longtable}
\usepackage{listings, lstautogobble}
\usepackage{booktabs}
\usepackage{threeparttable}
\usepackage[table,xcdraw]{xcolor}
\usepackage{wrapfig}
\usepackage{subfigure}
\usepackage{enumitem}
\usepackage{paralist}

\usepackage{amsthm}
\AtBeginEnvironment{thm}{\footnotesize}
\AtBeginEnvironment{defn}{\footnotesize}
\providecommand{\keywords}[1]{\textbf{\textit{Keywords---}} #1}

\usepackage{stackengine}
\newsavebox\mybox

\newtheorem{definition}{Definition}[section]

\newtheorem{theorem}{Theorem}[section]
\usepackage{hyperref}
\usepackage[framemethod=tikz]{mdframed}
\usepackage{enumitem}
\theoremstyle{definition}
\AtBeginEnvironment{thm}{\footnotesize}
\AtBeginEnvironment{defn}{\footnotesize}

{\begin{list}{}%
         {\setlength{\leftmargin}{#1}}%
         \item[]%
}
{\end{list}}

\newdateformat{monthyeardate}{%
  \monthname[\THEMONTH], \THEYEAR}
\begin{document}
\sloppy

\title{\Large{\textbf{Dynamic Dependability Analysis of Shuffle-exchange Networks using HOL Theorem Proving} }}
\author{
Yassmeen~Elderhalli, Osman~Hasan, and Sofi\`ene~Tahar\vspace*{2em}\\
Department of Electrical and Computer Engineering,\\
Concordia University, Montr\'eal, QC, Canada 
\vspace*{1em}\\
\{y\_elderh,o\_hasan,tahar\}@ece.concordia.ca 
 \vspace*{3em}\\
\textbf{TECHNICAL REPORT}\\
\date{October 2019}
}
\maketitle

\newpage
\begin{abstract}
Dynamic dependability models, such as dynamic fault trees (DFTs) and dynamic reliability block diagrams (DRBDs), are introduced to overcome the modeling limitations of traditional models. Recently, higher-order logic (HOL) formalizations of both models have been conducted, which allow the analysis of these models formally, within a theorem prover. In this report, we provide the formal dynamic dependability analysis of shuffle-exchange networks, which are multistage interconnection networks that are commonly used in multiprocessor systems. We use DFTs and DRBDs to model the terminal, broadcast and network reliability with dynamic spare gates and constructs in several generic versions. We verify generic expressions of probability of failure and reliability of these systems, which can be instantiated with any number of system components and failure rates to reason about the failure behavior of these networks.

\end{abstract}

\newpage
\section{Introduction}

Dependability describes the ability of a system to provide a trusted service~\cite{avizienis2004basic}. Dynamic dependability models, such as dynamic fault trees (DFTs)~\cite{DFT-survey} and dynamic reliability block diagrams~\cite{distefano2006new}, capture the dynamic failure and success dependencies, respectively, among system components, and hence are more suitable in modeling real-world systems. 
Recently, higher-order logic (HOL) theorem proving has been used in the formal analysis of both models algebraically~\cite{elderhalli2019methodology, elderhalli2019formally}, where generic expressions are formally verified that are independent of the failure distributions of system components. This ensures the soundness of the analysis, which is suitable for safety-critical systems. In this report, we use both formalizations in conducting the dynamic dependability analysis of the interconnection network of multiprocessor systems. 

\begin{figure}[hbtp]
\centering
\includegraphics[width=0.73\textwidth]{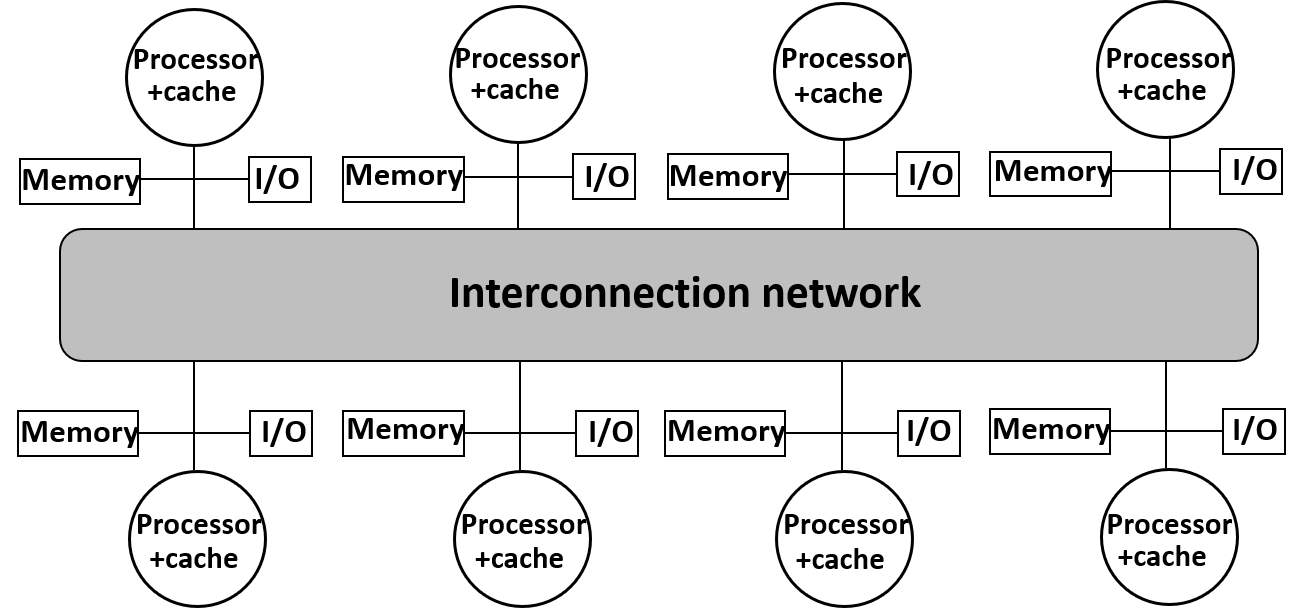}
\caption{Overview of Multiprocessor System Architecture}
\label{fig:multiprocessor}
\end{figure}

With the ongoing demands for intensive processing applications, multiprocessor systems represent one of the solutions that satisfies such demand. Nowadays, such systems are feasible due to their reduced cost and thus it is possible to have systems of hundreds of processors. Multiprocessor systems allow parallel computing, where tasks are executed in parallel with the possibility of interacting with one another when required. This parallel execution highly impacts the overall system performance, such as throughput. However, memory and I/O peripheral resources are shared among processors and thus an efficient data routing among system nodes is necessary to maintain high system performance, reliability and low cost. This is of a great importance, particularly with scientific applications, where a huge number of processors are used, i.e., large-scale multiprocessor systems~\cite{hennessy2011computer}. Therefore, a dedicated interconnection network is used to connect processors and memory modules, as depicted in Figure~\ref{fig:multiprocessor}~\cite{hennessy2011computer}.

The complexity of interconnection networks ranges from simple networks, such as time-shared bus to crossbar switching. The former has a negative impact on the system performance, while the latter has much higher cost as there exists a separate link between each pair of nodes in the systems. For example, for a system of $N$ nodes, i.e., $N$ inputs and $N$ outputs, it is required to have $N^2$ links or switching elements between each input and output. 

\begin{figure}[hbtp]
\centering
\includegraphics[scale=0.6]{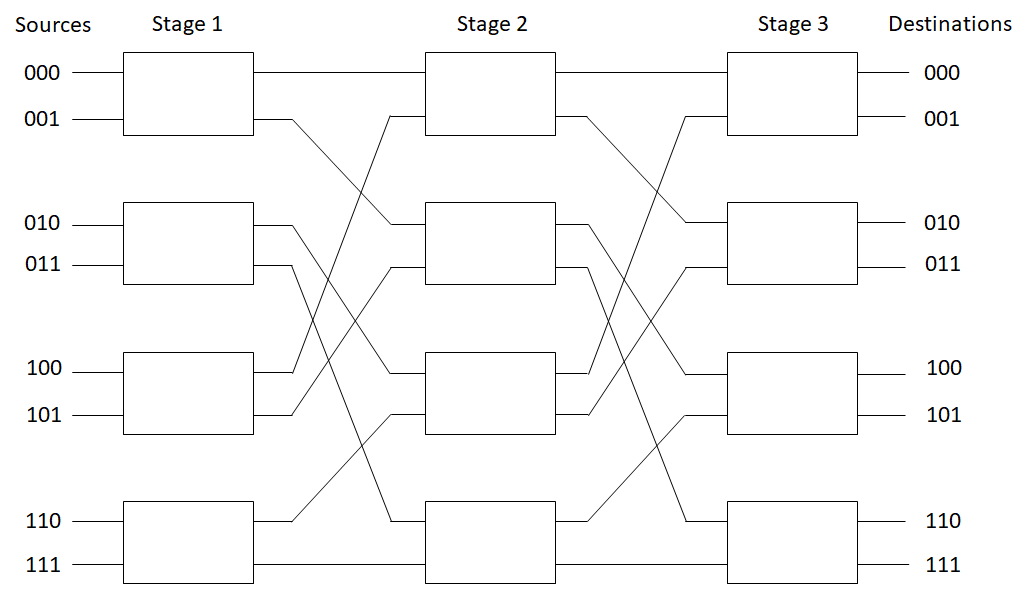}
\caption{An $8\times 8$ SEN}
\label{fig:SEN}
\end{figure}

Multistage interconnection networks (MINs) are introduced to reduce the number of required switching elements and hence, reduce the cost while providing better performance than shared-bus networks. The main idea of MINs is to have multiple small stages of crossbar switches that are connected between sources (inputs) and destinations (outputs), which results in a much reduced number of used switching elements. The number of paths available between each input and output determines the category of the MIN. A single-path MIN has only one path to route information between each source-destination pair. A shuffle-exchange network (SEN) is an example of such type of networks. Each stage has $log_{2}N$ switching elements, where $N$ is the number of inputs and outputs of the network. Usually the switching elements are of size $2\times 2$ to reduce the cost. The number of stages required to establish the single-path MIN is $Nlog_{2}N$, which is lower than crossbar networks. An $8 \times 8$ SEN is shown in Figure~\ref{fig:SEN}, where only a single path is available for each input-output pair. However, the reliability of single-path MINs and SENs depends on the switching elements and thus a fault in any of these switches cannot be tolerated. 

\begin{figure}[hbtp]
\centering
\includegraphics[scale=0.6]{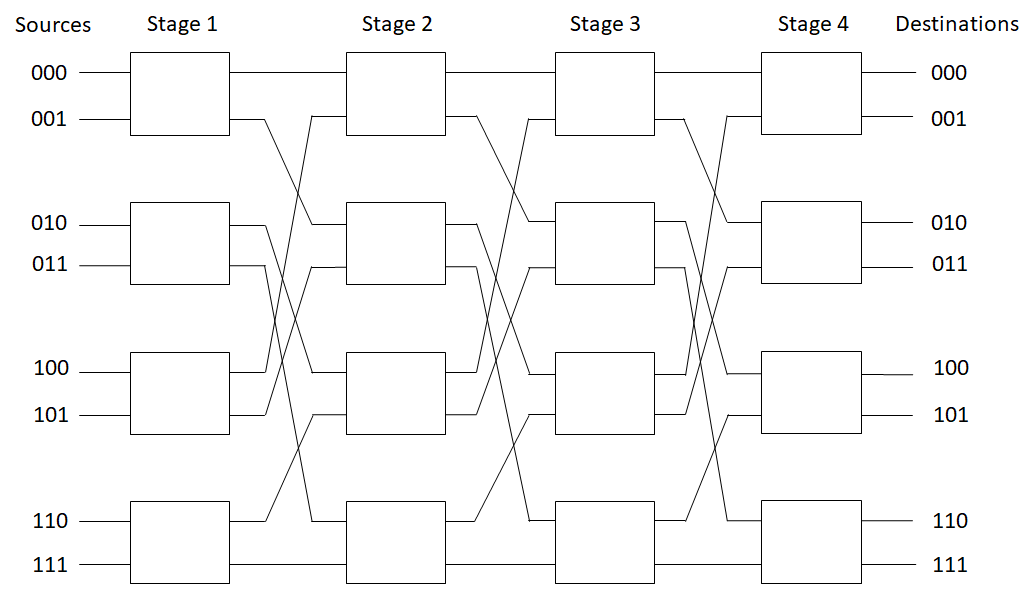}
\caption{An $8\times 8$ SEN+}
\label{fig:SEN+}
\end{figure}

Enhancing the reliability of MINs is of great importance in order to maintain high system performance. Therefore, redundant switching elements are used to ensure that the network is able to provide the required switching even after the failure of some of these elements~\cite{aggarwal2008reliability, kumar1988fault}. Multiple-path MINs are used to increase the fault tolerance and hence the network reliability. SEN+ is a SEN, where an additional stage is added to provide two paths between each input-output pair, as shown in Figure~\ref{fig:SEN+}. However, even with the additional path, the failure of some switches can lead to the failure of the connection in some situations. Spare parts have been used in~\cite{jeng1986fault} to replace switches after failure. However, the analysis was not conducted formally to ensure its correctness. 

Studying the reliability of SENs has been an active research area~\cite{yunus2015reliability,  bistouni2019determining, yunus2019evaluation, panda2018reliability}. 
The reliability of MINs are commonly analyzed using simulation or analytically. For example, in~\cite{gunawan2013reliability}, Monte Carlo simulation is used to analyze the reliability of SENs. However, as mentioned previously, simulation cannot provide accurate results due to its sampling based nature.  Although CTMCs can analytically solve the reliability of MINs~\cite{rajkumar2016review}, they cannot be used with large-scale systems since the state space grows exponentially with the increase in the number of system components. On the other hand, when the complexity of the network increases, reliability bounds provides estimate values for the MIN reliability~\cite{gunawan2008redundant, md2016reliability}.
RBDs have been also used in the analysis of MINs with single and multiple paths. For example, in~\cite{bistouni2014analyzing}, the reliability of SEN, SEN+ and SEN+2 (a SEN with two additional stages) is modeled using traditional RBDs. Generic expressions of success rates of the switching elements are provided analytically assuming that all these elements have the same failure rates. However, these generic expressions are not formally verified , which may raise questions about its accuracy. Furthermore, dynamic dependencies among system components, like warm spares, are not considered or modeled.   

Based on the previous discussion, accurate modeling and analysis  of these networks is necessary to capture the dynamic behavior as this will provide the design engineers with some measures that can help enhancing the performance of the entire multiprocessor system. To the best of our knowledge, dynamic dependability analysis using formal methods has not been used with MINs.  
Therefore, we propose to add spare switches to replace the critical ones after failure and conduct the analysis of MINs, particularly SENs using our formal dependability framework.

Since the reliability of MINs affects the performance of the overall multiprocessor system, it is required to accurately model and analyze their reliability. In this work, we use both DRBDs and DFTs to model the dynamic reliability of these networks, particularly SEN and SEN+, and conduct the analysis using our framework. In this work, we formally verify the terminal, broadcast and network reliability of SEN and SEN+ in HOL and provide generic expressions of reliability and probability of failure. It is worth noting that the formalization provided in this work uses the HOL theories (libraries) of DFT and DRBD, which have been developed in ~\cite{elderhalli2019formally, elderhalli2019methodology, elderhalli2019probabilistic} and can be accessed from ~\cite{DFT-code, Yassmeen-ICFEMcode}.

\section{Terminal Reliability Analysis of Shuffle-exchange Networks}

The terminal reliability is the reliability of the connection between a given source and destination, i.e., the probability of having a reliable connection between one source-destination pair. We analyze the terminal reliability of the SEN and SEN+ using both DFT and DRBD models.

\subsection{DFT Analysis of SEN and SEN+}

We model the sources of failure of both SEN and SEN+ using DFTs. We use $n$-ary gates, which enable verifying expressions of the probability of failure for generic number of system components. 

\begin{figure}[hbtp]
\centering
\includegraphics[scale=0.7]{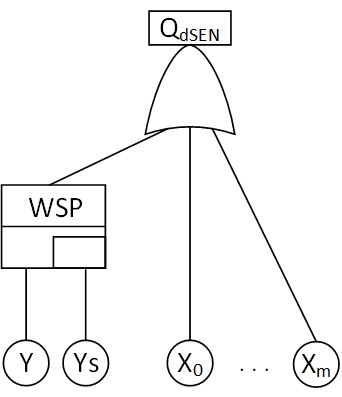}
\caption{DFT of SEN}
\label{fig:dft_sen}
\end{figure}

Figure~\ref{fig:dft_sen} shows the DFT model of the SEN system. Since SENs are single path MINs, the failure of any of the switches in the path between a given source and destination leads to losing the connection. Therefore, adding spare parts will lower the probability of failure. For illustration purposes, we use a spare part to replace the main switch $Y$ after failure. The DFT consists of an $n$-ary OR gate, which means that the failure of any of the switches, interrupts the connection between the source and the destination.

Since the top event is an $n$-ary OR gate, we need first to verify that the \texttt{DFT\_event} of the $n$-ary OR is equal to the union of the individual events, as:  \\
\vspace{50pt}

\begin{theorem}
\label{thm:n_OR_BIGUNION}
\emph{}\\
\mbox{\textup{\texttt{$\vdash\forall$ p X t s. FINITE s $\Rightarrow$}}}\\
\mbox{\textup{\texttt{~~(DFT\_event p (n\_OR (MAP X (SET\_TO\_LIST s))) t =}}}\\
\mbox{\textup{\texttt{~~~$\bigcup_{\texttt{i}\in \texttt{s}}$  \{rv\_to\_devent p X t i\})}}} 
\end{theorem}

\noindent where \texttt{s} is a set of numbers that has the indices of the system components. \texttt{X} is a group of random variables that represent the time-to-failure of the switches in the system. We need to recall that \texttt{n\_OR} accepts a list of random variables as an argument. Therefore, we create this list using \texttt{MAP X (SET\_TO\_LIST s)}. \texttt{rv\_to\_devent}, in Theorem~\ref{thm:n_OR_BIGUNION}, is similar to the \texttt{rv\_to\_event} of the DRBD, but it creates DFT events. It is defined as:

\begin{definition}
\textit{rv\_to\_devent}\\
\textup{\texttt{$\vdash\forall$ p X t. rv\_to\_devent p X t = ($\lambda$i. DFT\_event p (X i) t)}}
\end{definition}
 
This way, we can use this function to create a group of DFT events for a set of indexed random variables. 
Then, we verify the probability of the $n$-ary OR gate in a way similar to the probability of the DRBD parallel structure, which is defined as the union of events.

\begin{theorem}
\emph{}\\
\label{thm:prob_n_or}
\mbox{\textup{\texttt{$\vdash\forall$ p X t s. s $\neq$ \{\} $\wedge$ FINITE s }}}\\
\mbox{\textup{\texttt{~~indep\_sets p ($\lambda$i. \{rv\_to\_devent p X t i\}) s $\wedge$}}}\\
\mbox{\textup{\texttt{~~($\forall$ i. i $\in$ s $\Rightarrow$ rv\_gt0\_ninfinity [X i])  $\Rightarrow$}}}\\
\mbox{\textup{\texttt{~~(prob p (DFT\_event p (n\_OR (MAP X (SET\_TO\_LIST s))) t) =}}}\\
\mbox{\textup{\texttt{~~~1 - Normal ($\prod_{i\in s}$ (real (1 - F\textsubscript{X\textsubscript{i}}(t)))))}}}
 
\end{theorem}

In Theorem~\ref{thm:prob_n_or}, it is required that the set of indices, \texttt{s}, to be nonempty and to be finite, which is a realistic condition as in any system the number of components is finite. The last condition of Theorem~\ref{thm:term_dft_sSEN}, ensures that the random variables of \texttt{X} are greater than or equal to $0$ and not equal to $+\infty$, which is required to be able to use the CDF of the random variable as given in~\cite{elderhalli2019probabilistic}.

We express the structure function of the DFT of SEN as:

\begin{equation}
\begin{split}
\textup{\texttt{Q\textsubscript{dSEN\_Terminal} =}}& \textup{\texttt{ n\_OR (MAP ($\lambda$i. if i = 0 then WSP Y Ys\textsubscript{a} Ys\textsubscript{d}}}\\
&\textup{\texttt{~~~~~~~~~~~~~~~~~else X i) (SET\_TO\_LIST \{0\} $\cup$ L))}}
\end{split}
\end{equation}

We notice that the structure of the DFT is defined using the indices in \texttt{\{0\} $\cup$ L}. $0$ is the index of the spare gate and \texttt{L} has the indices of the rest of the switches in the system.

Finally, we verify the probability of failure of this top event as:

\begin{theorem}
\label{thm:term_dft_sSEN}
\emph{}\\
\textup{{\texttt{$\vdash\forall$ p X Y Ys\textsubscript{a} Ys\textsubscript{d} t L.}}}\\
\mbox{\textup{\texttt{~DISJOINT \{0\} L $\wedge$ FINITE L $\wedge$ L $\neq$ \{\} $\wedge$}}}\\
\mbox{\textup{\texttt{~indep\_sets p ($\lambda$i. \{event\_set [(DFT\_event p (WSP Y Ys\textsubscript{a} Ys\textsubscript{d}) t, 0)]}}}\\
\mbox{\textup{\texttt{~~~(rv\_to\_devent p X t) i\}) (\{0\} $\cup$ L) $\wedge$}}}\\
\mbox{\textup{{\texttt{~($\forall$ i. i $\in$ L $\Rightarrow$ rv\_gt0\_ninfinity [X i]) $\wedge$ }}}}\\
\mbox{\textup{{\texttt{~(prob p (DFT\_event p Q\textsubscript{dSEN\_Terminal} t) = }}}}\\
\mbox{\textup{\texttt{~~1-}}}\\
\mbox{\textup{\texttt{~~(1- prob p (DFT\_event p (WSP Y Ys\textsubscript{a} Ys\textsubscript{d}) t)) *}}}\\
\mbox{\textup{\texttt{~~~Normal ($\prod_{i\in L}$ (real(1-F\textsubscript{X\textsubscript{i}}(t))))}}} 
\end{theorem}

\noindent where \texttt{DISJOINT \{0\} L } ensures that the indices of the elements are unique. While \texttt{FINITE L $\wedge$ L $\neq$ \{\}} ascertain that set $L$, which has the indices, is finite and not empty. Finally, the independence of the events is  added using \texttt{indep\_sets}. Theorem~\ref{thm:term_dft_sSEN} can be further rewritten based on the probability of the spare gate~\cite{elderhalli2019probabilistic}. However, the required conditions of the latter should be satisfied, such as the continuity of the distributions. Since we need a group of indexed sets in \texttt{indep\_sets}, we define a function \texttt{event\_set} that accepts a list of pairs each of which is composed of a DFT event with its index. This function also accepts the remaining blocks of the DFT that have their indices embedded in a set (that can be generic of any size). 

\begin{figure}[!b]
\centering
\includegraphics[scale=0.70]{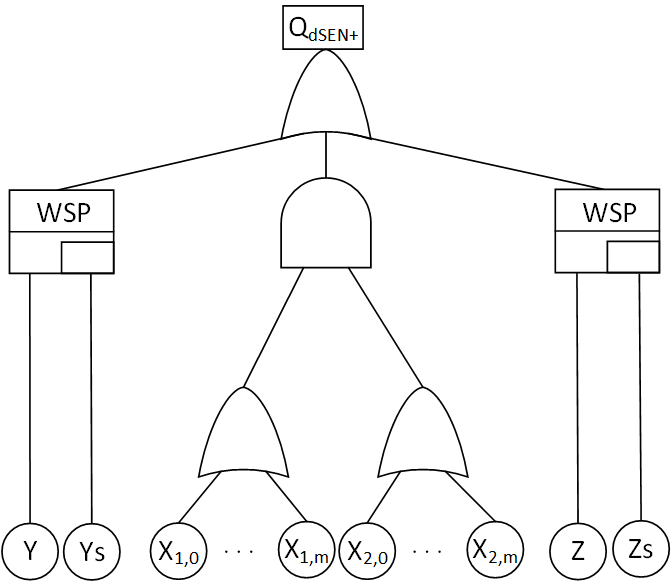}
\caption{DFT of SEN+ Terminal Connection}
\label{fig:DFT_SEN+}
\end{figure}

In SEN+, an additional path is added to increase the redundancy in the system. Therefore, for the connection between a given source and a destination to be broken, it is required that these two paths must be disconnected. The DFT of the SEN+ is shown in Figure~\ref{fig:DFT_SEN+}, where two spares are added to replace the main switches $Y$ and $Z$ after failure. Switch $Y$ is the input switch connected to the source and switch Z is connected to the destination. This DFT is composed of three levels of OR of AND of OR gates. Therefore, in order to verify the probability of the top event, we need first to verify that the \texttt{DFT\_event} of the $n$-ary AND gate is equal to the intersection of the input events. We formally verify this in HOL as:\\
\emph{}\\

\begin{theorem}
\label{thm:prob_n_AND}
\emph{}\\
\mbox{\textup{\texttt{$\vdash\forall$ p X t s.
          FINITE s  $\wedge$ s $\neq$ \{\} $\wedge$  0 $\leq$ t $\Rightarrow$}}}\\
\mbox{\textup{\texttt{~~(DFT\_event p }}}\\
\mbox{\textup{\texttt{~~~(n\_AND (MAP X (SET\_TO\_LIST s))) t = $\bigcap_{i\in s}$ \{rv\_to\_devent p X t i\})}}}
\end{theorem}

Then, we verify the probability of failure of the top event of the AND gate as:

\begin{theorem}
\emph{}\\
\label{thm:pron_n_and}
\mbox{\textup{\texttt{$\vdash\forall$ p X t s. FINITE s $\wedge$ s $\neq$ \{\} $\wedge$ 0 $\leq$ t $\wedge$ }}}\\
\mbox{\textup{\texttt{~~indep\_sets p ($\lambda$i. \{rv\_to\_devent p X t i\}) s}}}\\
\mbox{\textup{\texttt{~~($\forall$ i. i $\in$ s $\Rightarrow$ rv\_gt0\_ninfinity [X i]) $\Rightarrow$}}}\\
\mbox{\textup{\texttt{~~(prob p}}}\\
\mbox{\textup{\texttt{~~~(DFT\_event p}}}\\
\mbox{\textup{\texttt{~~~~(n\_AND (MAP X (SET\_TO\_LIST s))) t) =}}}\\
\mbox{\textup{\texttt{~~~Normal ($\prod_{i\in s}$ (real (F\textsubscript{X\textsubscript{i}}(t)))))}}}
\end{theorem}

The first three conditions are needed to be able to use Theorem~\ref{thm:prob_n_AND}, while \texttt{indep\_sets} ensures the independence of the events. 

We use Theorems~\ref{thm:prob_n_or} and~\ref{thm:pron_n_and} to verify the probability of OR of AND of OR, which is required for the probability of the top event. We express the top event of the DFT of Figure~\ref{fig:DFT_SEN+}, \texttt{Q\textsubscript{dSEN+}} as: 

\begin{equation}
\begin{split}
\textup{\texttt{Q\textsubscript{dSEN+\_Terminal}}} =& {\textup{\texttt{ n\_OR (MAP ($\lambda$i. if i = 0 then WSP Y Ys\textsubscript{a} Ys\textsubscript{d}}}}\\
&{{\textup{\texttt{~~~~~~~~~~~~~~~~~else if i = 1 then}}}}\\
&{\textup{\texttt{~~~~~~~~~~~~~~~~~~~\big((n\_OR (MAP X (SET\_TO\_LIST L1))) $\cdot$}}}\\
&\textup{\texttt{~~~~~~~~~~~~~~~~~~~~(n\_OR (MAP X (SET\_TO\_LIST L2)))\big)}}\\
&{{\textup{\texttt{~~~~~~~~~~~~~~~~~else WSP Z Zs\textsubscript{a} Zs\textsubscript{d}) (SET\_TO\_LIST \{0; 1; 2\}))}}}}
\end{split}
\end{equation}
\noindent where $\{0;1;2\}$ indicates that the OR gate has three inputs with indices $0$ for the first spare, $1$ for the AND of ORs, and $2$ for the second spare. $L1$ and $L2$ has the indices of the switches in the two redundant paths (for the two lower ORs).

The DFT top event can be expressed using union and intersection of events, which can be quite useful in reusing the existing theorems of probability of union of intersections and intersection of unions. We verify this relationship as:

\begin{theorem}
\label{thm:dft_SEN_terminal_n_OR}
\emph{}\\
\mbox{\textup{\texttt{$\vdash\forall$ p Y Ys\textsubscript{a} Ys\textsubscript{d} Z Zs\textsubscript{a} Zs\textsubscript{d} X L1 L2 t.}}}\\
\mbox{\textup{\texttt{~~FINITE L1 $\wedge$ FINITE L2 $\wedge$}}}\\
\mbox{\textup{\texttt{~~disjoint\_family\_on (ind\_set [\{0\}; L1; L2; \{3\}]) \{0; 1; 2; 3\} $\Rightarrow$}}}\\
\mbox{\textup{\texttt{~~(DFT\_event p (Q\textsubscript{dSEN+\_Terminal}) t =}}}\\

\noindent \mbox{\textup{\texttt{~~~$\bigcup$}}}\\
\mbox{\textup{\texttt{~~~~~\{$\bigcap$}}}\\
\mbox{\textup{\texttt{~~~~~~~\{$\bigcup$}}}\\
\mbox{\textup{\texttt{~~~~~~~~~\{event\_set}}}\\
\mbox{\textup{\texttt{~~~~~~~~~~~[(DFT\_event p (WSP Y Ys\textsubscript{a} Ys\textsubscript{d}) t,0);}}}\\
\mbox{\textup{\texttt{~~~~~~~~~~~~(DFT\_event p (WSP Z Zs\textsubscript{a} Zs\textsubscript{d}) t,3)]}}}\\
\mbox{\textup{\texttt{~~~~~~~~~(rv\_to\_devent p X t) i |}}}\\
\mbox{\textup{\texttt{~~~~~~~~i $\in$ ind\_set [\{0\}; L1; L2; \{3\}] a\} | a |}}}\\
\mbox{\textup{\texttt{~~~~~~a $\in$ ind\_set [\{0\}; \{1; 2\}; \{3\}] j\} | j |}}}\\
\mbox{\textup{\texttt{~~~~~j $\in$ \{0; 1; 2\}\})}}}
\end{theorem}

Finally, we verify the probability of failure of \texttt{Q\textsubscript{dSEN+}}:

\begin{theorem}

\emph{}\\
\textup{{\texttt{$\vdash\forall$ p X Y Ys\textsubscript{a} Ys\textsubscript{d} Z Zs\textsubscript{a} Zs\textsubscript{d} t L1 L2. 0 $\leq$ t $\wedge$}}}\\
\mbox{\textup{{\texttt{~SEN\_set\_req p L1 L2 (ind\_set [\{0\}; L1; L2; \{3\}])}}}}\\
\mbox{\textup{{\texttt{~~~(ind\_set [\{0\}; \{1; 2\}; \{3\}]) \{0; 1; 2\}}}}}\\
\mbox{\textup{{\texttt{~~~(event\_set [(DFT\_event p (WSP Y Ys\textsubscript{a} Ys\textsubscript{d}) t,0);}}}}\\
\mbox{\textup{{\texttt{~~~~~~~~~~~~~~~(DFT\_event p (WSP Z Zs\textsubscript{a} Zs\textsubscript{d}) t,3)]}}}}\\
\mbox{\textup{\texttt{~~~~~~~~~(rv\_to\_devent p X t)) $\wedge$}}}\\
\mbox{\textup{\texttt{~($\forall$ i. i $\in$ (L1 $\cup$ L2) $\Rightarrow$ rv\_gt0\_ninfinity [X i]) $\Rightarrow$}}}\\
\mbox{\textup{{\texttt{~(prob p (DFT\_event p
           Q\textsubscript{dSEN+\_Terminal} t) =}}}}\\
\mbox{\textup{\texttt{~~1 -}}}\\
\mbox{\textup{\texttt{~~~(1 -}}}\\
\mbox{\textup{\texttt{~~~~prob p (DFT\_event p (WSP Y Ys\textsubscript{a} Ys\textsubscript{d}) t)) *}}}\\
\mbox{\textup{\texttt{~~~~(Normal}}}\\
\mbox{\textup{\texttt{~~~~~(1 -}}}\\
\mbox{\textup{\texttt{~~~~~~~(1 - $\prod_{i\in L1}$ (real (1 - F\textsubscript{X\textsubscript{i}}(t)))) *}}}\\
\mbox{\textup{\texttt{~~~~~~~(1 - $\prod_{i\in  L2}$ (real (1 - F\textsubscript{X\textsubscript{i}}(t))))) *}}}\\
\mbox{\textup{\texttt{~~~(1 - prob p (DFT\_event p (WSP Z Zs\textsubscript{a} Zs\textsubscript{d}) t))))}}}
\end{theorem}

\noindent where \texttt{SEN\_set\_req} ensures the required conditions of the input sets including that the sets are finite and nonempty. It also ensures the independence of the input events over the probability space. We also define \texttt{ind\_set} that accepts a list of sets and returns a group of indexed sets.
This is required to be able to create the hierarchy of the DFT using sets.
 
\begin{figure}[!t]
\centering
\vspace{10pt}
\includegraphics[scale=1]{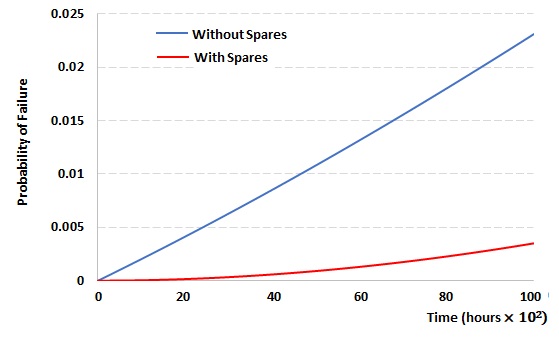}
\caption{Probability of Failure of the Terminal Connection of a $128\times 128$ SEN+ with and without Spares}
\label{fig:dft_term_plot}
\end{figure}

In order to use the above generic probability of failure expressions on a concrete instance of SEN+, we evaluate in MATLAB~\cite{MATLAB} the probability of failure of the terminal connection of a $128\times 128$ SEN+, where each OR gate of the first level of Figure~\ref{fig:DFT_SEN+} has 6 inputs. We assume  that the failure rate of each switching element is  $1\times 10^{-5}$. We evaluate the probability of failure for the SEN+ system without and with spare parts with a dormancy factor of 0.1, as shown in Figure~\ref{fig:dft_term_plot}. This result shows that considering the spares in the analysis leads to having more reliable and realistic system than the traditional FTs.

\subsection{DRBD Analysis of SEN and SEN+}

\begin{figure}[hbtp]
\centering
\includegraphics[scale=1]{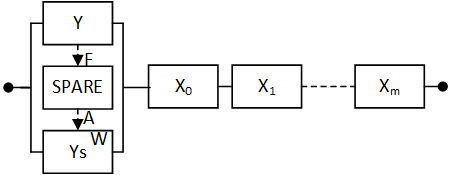}
\caption{DRBD of SEN}
\label{fig:term_sen_DRBD}
\end{figure}

For SENs (single-path MIN), the terminal reliability is modeled as a series RBD. For illustration purposes, we use a spare part to replace the first input switch, and thus increase the reliability. The DRBD of the modified SEN is shown in Figure~\ref{fig:term_sen_DRBD}, where $Y$ is the main switch that will be replaced by $Ys$ after failure and the series structure has $m+1$ elements.

Using the proposed DRBD algebra in~\cite{elderhalli2019formally}, we express the structure function of the SEN DRBD as:

\begin{equation}
\begin{split}
\textup{\texttt{Q\textsubscript{SEN\_Terminal} = }}& \textup{\texttt{nR\_AND ($\lambda$i. if i = 0 then R\_WSP Y Ys\textsubscript{a} Ys\textsubscript{d}}}\\
&\textup{\texttt{~~~~~~~~~~~~~else X i) \{0\} $\cup$ L}}
\end{split}
\end{equation}

\noindent where $X$ is a group of indexed time-to-failure functions that represent the blocks of the series structure and $L$ is a set with their indices. $L$ can be instantiated with any group of numbers, which makes this function generic to represent the reliability model of any SEN with any size.

Then, we verify that the \texttt{DRBD\_event} of \texttt{Q\textsubscript{SEN}} can be represented using the series parallel structures as:

\begin{theorem}
\label{thm:sSEN_nR_AND}
\emph{}\\
\textup{{\texttt{$\vdash\forall$ p X Y Ys\textsubscript{a} Ys\textsubscript{d} t L.}}}\\
\mbox{\textup{{\texttt{~DISJOINT \{0\} L $\wedge$ FINITE L $\wedge$ L $\neq$ \{\} $\Rightarrow$}}}}\\
\mbox{\textup{{\texttt{~(DRBD\_event p Q\textsubscript{SEN\_Terminal} t =}}}}\\
\mbox{\textup{{\texttt{~~DRBD\_series }}}}\\
\mbox{\textup{{\texttt{~~~($\lambda$i. event\_set}}}}\\
\mbox{\textup{{\texttt{~~~~[(DRBD\_event p (R\_WSP Y Ys\textsubscript{a} Ys\textsubscript{d}) t,0)]}}}}\\
\mbox{\textup{{\texttt{~~~~~(rv\_to\_event p X t) i) (\{0\} $\cup$ L) )}}}}
\end{theorem}

\noindent where \texttt{DISJOINT} ensures that all sets are disjoint. We use \texttt{event\_set} and \texttt{ind\_set} to create the events, similar to the DFTs. Since we are dealing with a series structure, we only need to specify the heirarchy of the architecture in one direction using $\{0\}\cup L$. We verify Theorem \ref{thm:sSEN_nR_AND} using the relationship between \texttt{nR\_AND} and \texttt{DRBD\_series} verified in~\cite{elderhalli2019formally} and some set-related theorems. 

Based on Theorem \ref{thm:sSEN_nR_AND}, we verify a generic expression for the reliability of the SEN system:
\begin{theorem}
\label{thm:term_Rel_sSEN}
\emph{}\\
\textup{{\texttt{$\vdash\forall$ p X Y Ys\textsubscript{a} Ys\textsubscript{d} t L.}}}\\
\mbox{\textup{{\texttt{~DISJOINT \{0\} L $\wedge$ FINITE L $\wedge$ L $\neq$ \{\} $\wedge$}}}}\\
\mbox{\textup{\texttt{~indep\_sets p ($\lambda$i. \{event\_set [(DRBD\_event p (R\_WSP Y Ys\textsubscript{a} Ys\textsubscript{d}) t, 0)]}}}\\
\mbox{\textup{\texttt{~~~(rv\_to\_event p X t) i\}) (\{0\} $\cup$ L)$\Rightarrow$}}}\\
\mbox{\textup{{\texttt{~(prob p (DRBD\_event p Q\textsubscript{SEN\_Terminal} t) =}}}}\\
\mbox{\textup{{\texttt{~~Rel p (R\_WSP Y Ys\textsubscript{a} Ys\textsubscript{d}) t * Normal ($\prod_{l\in L}$ (real (Rel p (X l) t)))) }}}}
\end{theorem}

\begin{figure}[hbtp]
\centering
\vspace{10pt}
\includegraphics[width= 0.7\textwidth]{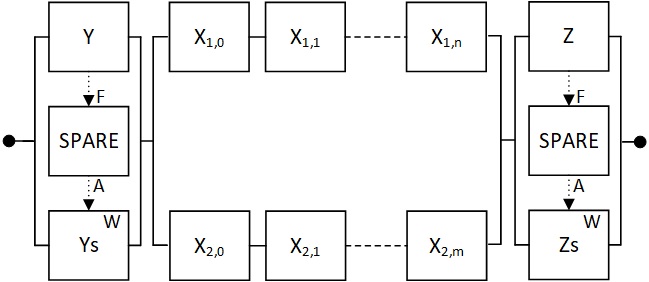}
\caption{Terminal Reliability DRBD of SEN+}
\label{fig:term_sen+_drbd}
\end{figure}

In a similar manner, the SEN+ is modeled as a series-parallel-series structure. To further enhance the reliability, we use spare constructs as shown in Figure~\ref{fig:term_sen+_drbd}, where $Y$ and $Z$ are the main single switches that are connected to the source and destination with their spares $Ys$ and $Zs$, respectively. The parallel structure in the middle represents the reliability model of the two alternative paths between the source and the destination. Therefore, this DRBD consists of a series of two spare constructs and one parallel structure that consists of two series structures.

Using our DRBD operators, we formally express the structure function of this DRBD as:
\begin{equation}
\begin{split}
\small{\textup{\texttt{Q\textsubscript{SEN+\_Terminal}}}} = &\small{\textup{\texttt{ nR\_AND ($\lambda$i. if i = 0 then R\_WSP Y Ys\textsubscript{a} Ys\textsubscript{d}}}}\\
&{\small{\textup{\texttt{~~~~~~~~~~~~~~else if i = 1 then \big((nR\_AND X L1) + (nR\_AND X L2)\big)}}}}\\
&{\small{\textup{\texttt{~~~~~~~~~~~~~~else R\_WSP Z Zs\textsubscript{a} Zs\textsubscript{d}) \{0; 1; 2\}}}}}
\end{split}
\end{equation}

Thus, the outer series structure is expressed using the \texttt{nR\_AND} operator over the set $\{0;1;2\}$ as this structure has three different structures; i.e., two spare constructs and one parallel structure. 
In order to re-utilize the verified expressions of reliability, it is required to express this DRBD using the series and parallel structures. Therefore, we verify that the DRBD event of the \texttt{Q\textsubscript{SEN+}} is equal to a nested series-parallel-series structure as:

\begin{theorem}
\label{thm:SEN_nR_AND}
\emph{}\\
\textup{{\texttt{$\vdash\forall$ p X Y Ys\textsubscript{a} Ys\textsubscript{d} Z Zs\textsubscript{a} Zs\textsubscript{d} t L1 L2.}}}\\
\mbox{\textup{{\texttt{~disjoint\_family\_on (ind\_set [\{0; 3\}; L1; L2]) \{0;1;2\} $\wedge$ }}}}\\
\mbox{\textup{\texttt{~FINITE L1 $\wedge$ FINITE L2 $\wedge$ L1 $\neq$ \{\} $\wedge$ L2 $\neq$ \{\} $\Rightarrow$}}}\\
\mbox{\textup{{\texttt{~(DRBD\_event p Q\textsubscript{SEN+\_Terminal} t =}}}}\\
\mbox{\textup{{\texttt{~~DRBD\_series ($\lambda$j.}}}}\\
\mbox{\textup{{\texttt{~~~~DRBD\_parallel ($\lambda$a.}}}}\\
\mbox{\textup{{\texttt{~~~~~~DRBD\_series ($\lambda$i.}}}}\\
\mbox{\textup{{\texttt{~~~~~~~~event\_set}}}}\\
\mbox{\textup{{\texttt{~~~~~~~~~[(DRBD\_event p (R\_WSP Y Ys\textsubscript{a} Ys\textsubscript{d}) t,0);}}}}\\
\mbox{\textup{{\texttt{~~~~~~~~~~(DRBD\_event p (R\_WSP Z Zs\textsubscript{a} Zs\textsubscript{d}) t,3)]}}}}\\
\mbox{\textup{{\texttt{~~~~~~~~~(rv\_to\_event p X t) i)}}}}\\
\mbox{\textup{{\texttt{~~~~~~~ind\_set [\{0\}; L1; L2; \{3\}] a))}}}}\\
\mbox{\textup{{\texttt{~~~~~(ind\_set [\{0\}; \{1; 2\}; \{3\}] j)) \{0; 1; 2\})}}}}
\end{theorem}

\noindent where \texttt{disjoint\_family\_on (ind\_set [\{0; 3\}; L1; L2]) \{0;1;2\}} ensures that the sets $\{0;3\}$, $L1$ and $L2$ are disjoint, i.e., each switch has a unique index. Since we are dealing with a series-parallel-series structure, we need three sets to identify the hierarchy of this nested structure. Set $\{0;1;2\}$ in Theorem \ref{thm:SEN_nR_AND} indicates that the outer series structure has three elements, i.e., three parallel structures. \texttt{ind\_set [\{0\}; \{1;2\}; \{3\}]} indicates that the first parallel structure has only one series structure with index $0$, the second parallel structure has two series structures with indices $1$ and $2$, and the third parallel structure has only one series structure with index $3$. Finally, \texttt{ind\_set [\{0\}; L1; L2; \{3\}]} implies that the first series structure has only one element with index $0$, the second and third series structures have an arbitrary number of blocks indexed by $L1$ and $L2$. The last series structure has one element with index $3$. We verify Theorem~\ref{thm:SEN_nR_AND} using the relationship between the event of \texttt{nR\_AND} and the \texttt{DRBD\_series} and the equivalence of the event of the OR with the union of events besides some set-related theorems. 

Based on Theorem \ref{thm:SEN_nR_AND}, we verify a generic expression for the reliability of the SEN+ system:
\begin{theorem}
\label{thm:Rel_SEN}
\emph{}\\
\textup{{\texttt{$\vdash\forall$ p X Y Ys\textsubscript{a} Ys\textsubscript{d} Z Zs\textsubscript{a} Zs\textsubscript{d} t L1 L2.}}}\\
\mbox{\textup{{\texttt{~SEN\_set\_req p L1 L2 (ind\_set [\{0\}; L1; L2; \{3\}])}}}}\\
\mbox{\textup{{\texttt{~~~(ind\_set [\{0\}; \{1; 2\}; \{3\}]) \{0; 1; 2\}}}}}\\
\mbox{\textup{{\texttt{~~~(event\_set [(DRBD\_event p (R\_WSP Y Ys\textsubscript{a} Ys\textsubscript{d}) t,0);}}}}\\
\mbox{\textup{{\texttt{~~~~~~~~~~~~~~~(DRBD\_event p (R\_WSP Z Zs\textsubscript{a} Zs\textsubscript{d}) t,3)]}}}}\\
\mbox{\textup{\texttt{~~~~~~~~~~(rv\_to\_event p X t)) $\Rightarrow$}}}\\
\mbox{\textup{{\texttt{~(prob p (DRBD\_event p Q\textsubscript{SEN\_Terminal} t) =}}}}\\
\mbox{\textup{{\texttt{~~Rel p (R\_WSP Y Ys\textsubscript{a} Ys\textsubscript{d}) t * Rel p (R\_WSP Z Zs\textsubscript{a} Zs\textsubscript{d}) t *}}}}\\
\mbox{\textup{{\texttt{~~(1 - }}}}\\
\mbox{\textup{\texttt{~~~~(1 - Normal ($\prod_{l\in L1}$ (real (Rel p (X l) t)))) *}}}\\
\mbox{\textup{\texttt{~~~~(1 - Normal ($\prod_{l\in L2}$ (real (Rel p (X l) t))))))}}}
\end{theorem}

\noindent where \texttt{SEN\_set\_req} is the same function that we use with DFTs. We first rewrite the goal using Theorem \ref{thm:SEN_nR_AND}, then we use the reliability of the series-parallel-series to verify  the final expression. The reliability of the spare constructs can be further rewritten using the probability of the spare construct verified in ~\cite{elderhalli2019formally} given that the required conditions are ensured, such as the continuity of the CDFs. It can be noticed that the DRBD and the DFT models possess the same hierarchy represented by the sets of indices, which  makes it easy to be used when going from one model to the other.

\begin{figure}[hbtp]
\centering
\includegraphics[scale=1]{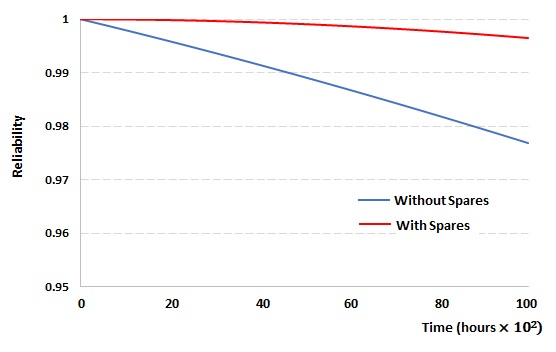}
\caption{Terminal Reliability of $128\times 128$ SEN+ with and without Spares}
\label{fig:drbd_term_plot}
\end{figure}

Similar to the DFT analysis, we evaluate the terminal reliability of a $128\times 128$ SEN+, where each inner series structure of Figure~\ref{fig:term_sen+_drbd} has 6 blocks. We assume that the failure rate of each switching element is  $1\times 10^{-5}$. We evaluate the reliability for the SEN+ system without and with spare parts with a dormancy factor of 0.1, as shown in Figure~\ref{fig:drbd_term_plot}.  

\section{Broadcast Reliability Analysis of Shuffle-exchange Networks}

The broadcast reliability represents the probability of having a working connection between one source and all destinations. This is required when one of the processors in the system needs to transmit information to all destinations in the network. We present in this section, the broadcast reliability of the SEN and SEN+ using both DFT and DRBD models. 

\subsection{DFT Analysis of SEN and SEN+}

Since in SENs there exists a single path between each source and destination, it is required to have a successful transmission through all these paths for a proper broadcast. Therefore, the DFT can be modeled using an OR gate. We further lower the probability of failure by adding an additional spare gate, as shown in Figure~\ref{fig:dft_sen}. However, the number of DFT inputs, which represent the switches, varies between the terminal and broadcast reliability models. For example, consider an $8\times 8$ SEN. The number of inputs for the terminal DFT is $3$, i.e., $log_{2} 8$, while the broadcast DFT requires seven inputs, i.e., $\sum_{i=1}^{log_{2}8}(\frac{8}{2^i})$~\cite{bistouni2014analyzing}. Therefore, we can also use Theorem~\ref{thm:term_dft_sSEN} for the broadcast, since this theorem is verified for any number of system blocks with their indices in the set $s$ . This highlights the importance of having generic verified expressions for any number of system blocks, which enables the re-utilization of the theorems in different contexts. 

\begin{figure}[hbtp]
\centering
\includegraphics[scale = 0.7]{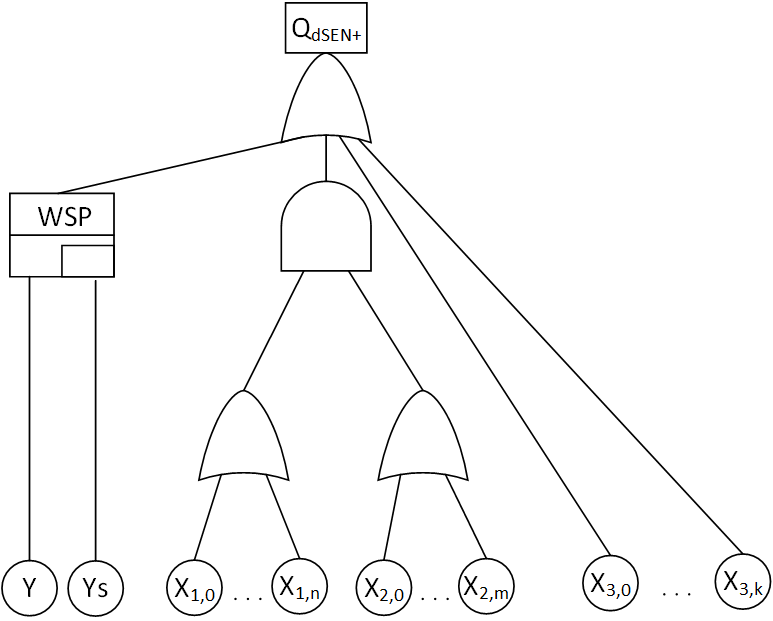}
\caption{DFT of Broadcast SEN+}
\label{fig:broadcast_dft_sen+}
\end{figure}

The DFT model of the broadcast SEN+ is shown in Figure~\ref{fig:broadcast_dft_sen+}. Its top event is modeled using an OR gate that is connected to a spare gate for the input switch, AND of OR to model the two alternative paths and finally, the rest of the destination switches in order to have a proper broadcast transmission. 

We formally express the structure function of the top event as:

\begin{equation}
\begin{split}
\textup{\texttt{Q\textsubscript{dSEN+\_Broadcast}}} =& {\textup{\texttt{ n\_OR (MAP ($\lambda$i. if i = 0 then WSP Y Ys\textsubscript{a} Ys\textsubscript{d}}}}\\
&{{\textup{\texttt{~~~~~~~~~~~~~~else if i = 1 then}}}}\\
&{\textup{\texttt{~~~~~~~~~~~~~~~~\big((n\_OR (MAP X (SET\_TO\_LIST L1))) $\cdot$}}}\\
&\textup{\texttt{~~~~~~~~~~~~~~~~~(n\_OR (MAP X (SET\_TO\_LIST L2)))\big)}}\\
&{{\textup{\texttt{~~~~~~~~~~~~~~else (n\_OR (MAP X (SET\_TO\_LIST L3))))}}}}\\
&\textup{\texttt{~~~~~~~~~~(SET\_TO\_LIST \{0; 1; 2\}))}}
\end{split}
\end{equation}

The hierarchy of the DFT is divided using the sets of indices. We need to recall that \texttt{MAP X (SET\_TO\_LIST L1)}, \texttt{MAP X (SET\_TO\_LIST L2)} and \texttt{MAP X (SET\_TO\_LIST L3)} are used to create the lists of the group of random variables for the $n$-ary gates. $L1$ and $L2$ has the indices of the switches in the two alternative paths, i.e., the inputs of the two lower OR gates in the DFT of Figure~\ref{fig:broadcast_dft_sen+}, while $L3$ has the indices of the remaining inputs of the top OR gate. The set $\{0;1;2\}$ indicates that the top OR gate has three inputs, which is similar to the terminal DFT model. 

We use this structure function to verify the probability of failure of the top event:

\begin{theorem}
\label{thm:brodcast_dft_SEN}
\emph{}\\
\textup{{\texttt{$\vdash\forall$ p X Y Ys\textsubscript{a} Ys\textsubscript{d} t L1 L2 L3 s.}}}\\
\mbox{\textup{{\texttt{~SEN\_broad\_set\_req p L1 L2 L3 (ind\_set [\{0\}; L1; L2; L3])}}}}\\
\mbox{\textup{{\texttt{~~~(ind\_set [\{0\}; \{1; 2\}; \{3\}]) \{0; 1; 2\}}}}}\\
\mbox{\textup{{\texttt{~~~(event\_set [(DFT\_event p (WSP Y Ys\textsubscript{a} Ys\textsubscript{d}) t,0);}}}}\\
\mbox{\textup{{\texttt{~~~~~(rv\_to\_devent p X t)) $\wedge$ 0 $\leq$ t $\wedge$ }}}}\\
\mbox{\textup{\texttt{~($\forall$ i. i $\in$ (L1 $\cup$ L2 $\cup$ L3) $\Rightarrow$ rv\_gt0\_ninfinity [X i]) $\Rightarrow$}}}\\
\mbox{\textup{{\texttt{~(prob p (DFT\_event p Q\textsubscript{dSEN+\_Broadcast} t) =}}}}\\
\mbox{\textup{\texttt{~~1 -}}}\\
\mbox{\textup{\texttt{~~~(1 -}}}\\
\mbox{\textup{\texttt{~~~~prob p (DFT\_event p (WSP Y Ys\textsubscript{a} Ys\textsubscript{d} t)) *}}}\\
\mbox{\textup{\texttt{~~~~(Normal}}}\\
\mbox{\textup{\texttt{~~~~~(1 -}}}\\
\mbox{\textup{\texttt{~~~~~~~(1 - $\prod_{i\in L1}$ (real (1 - F\textsubscript{X\textsubscript{i}}(t)))) *}}}\\
\mbox{\textup{\texttt{~~~~~~~(1 - $\prod_{i\in  L2}$ (real (1 - F\textsubscript{X\textsubscript{i}}(t))))) *}}}\\
\mbox{\textup{\texttt{~~~Normal ($\prod_{i\in L3}$ (real (1 - F\textsubscript{X\textsubscript{i}}(t))))))}}}
\end{theorem}
 
\noindent where \texttt{SEN\_broad\_set\_req} ascertains the conditions required for the sets such as finiteness. It also ensures the independence of the events.

\begin{figure}[hbtp]
\centering
\includegraphics[scale=1]{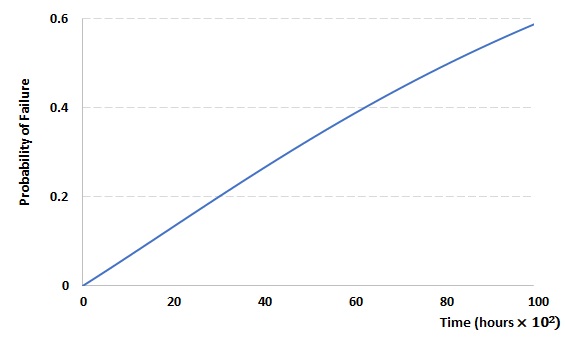}
\caption{Probability of Failure of the Broadcast of a $128\times 128$ SEN+}
\label{fig:dft_broad_plot}
\end{figure}

Figure~\ref{fig:dft_broad_plot} shows the evaluation results of the probability of failure of the DFT of Figure~\ref{fig:broadcast_dft_sen+} for a $128 \times 128$ SEN+. This SEN+ has 63 inputs for each first level OR gate and the top level OR gate has 66 inputs. As with the terminal SEN+, we assume that the failure rate of each switching element is  $1\times 10^{-5}$ with a dormancy factor of $0.1$.  

\subsection{DRBD Analysis of SEN and SEN+}

Similar to the DFT SEN broadcast model, we can use the model in Figure~\ref{fig:term_sen_DRBD}. However, as mentioned previously, the number of the blocks is different. Therefore, we can also use Theorem~\ref{thm:term_Rel_sSEN} for the broadcast reliability, since this theorem is verified for any number of system blocks using set $s$.

\begin{figure}[hbtp]
\centering
\includegraphics[width=\textwidth]{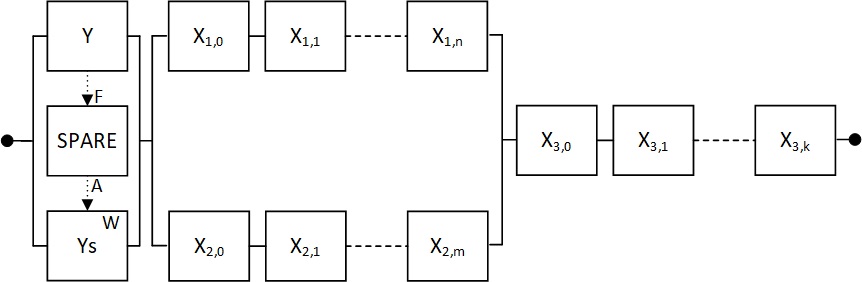}
\caption{Broadcast DRBD model of SEN+}
\label{fig:broadcast_rel_sen+}
\end{figure}

The DRBD of the SEN+ is depicted in Figure~\ref{fig:broadcast_rel_sen+}. The first block (with the spare) represents the input switch that is connected directly to the source. The failure of this switch will interrupt the broadcast transmission. Therefore, we add a spare part to replace it after failure. The series structure on the right side of the figure models the switches of all destinations, as they are all receiving the transmission. Finally, the parallel-series structure in the middle, represents the two alternative paths that are available for each broadcast transmission. For example, for the SEN+ shown in Figure~\ref{fig:SEN+}, the number of switches connected to the destinations are four, while each one of the alternative paths has three switches.   

In order to formally verify the reliability of the broadcast of the SEN+, we first express it using our operators as:

\begin{equation}
\begin{split}
\textup{\texttt{Q\textsubscript{SEN+\_Broadcast}}} =& {\textup{\texttt{ nR\_AND ($\lambda$i. if i = 0 then R\_WSP Y Ys\textsubscript{a} Ys\textsubscript{d}}}}\\
&{{\textup{\texttt{~~~~~~~~~~~~~~else if i = 1 then \big((nR\_AND X L1) $+$ }}}}\\
&{{\textup{\texttt{~~~~~~~~~~~~~~~~~~~~~~~~~~~~~~~~~~(nR\_AND X L2)\big)}}}}\\
&{{\textup{\texttt{~~~~~~~~~~~~~~else (nR\_AND X L3)) (\{0; 1\; 2\})}}}}
\end{split}
\end{equation}

\noindent where $L1$ and $L2$ are the sets that have the indices of the inner series structures of the parallel-series structure in the middle. The set $\{0; 1; 2\}$ indicates that the outer series structure  consists of three main components. The first spare construct has index $0$, while the parallel-series structure has index $1$. Finally, the series structure on the left side of Figure~\ref{fig:broadcast_rel_sen+} has index $2$, and $L3$ has the indices of the blocks in this series structure. We verify the reliability of this DRBD as:

\begin{theorem}
\label{thm:brodcast_Rel_SEN}
\emph{}\\
\textup{{\texttt{$\vdash\forall$ p X Y Ys\textsubscript{a} Ys\textsubscript{d} t L1 L2 L3.}}}\\
\mbox{\textup{{\texttt{~SEN\_broad\_set\_req p L1 L2 (ind\_set [\{0\}; L1; L2; L3])}}}}\\
\mbox{\textup{{\texttt{~~~(ind\_set [\{0\}; \{1; 2\}; \{3\}]) \{0; 1; 2\}}}}}\\
\mbox{\textup{{\texttt{~~~(event\_set [(DRBD\_event p (R\_WSP Y Ys\textsubscript{a} Ys\textsubscript{d}) t,0);}}}}\\
\mbox{\textup{{\texttt{~~~~~(rv\_to\_event p X t)) $\Rightarrow$}}}}\\
\mbox{\textup{{\texttt{~(prob p (DRBD\_event p Q\textsubscript{SEN+\_Broadcast} t) =}}}}\\
\mbox{\textup{{\texttt{~~Rel p (R\_WSP Y Ys\textsubscript{a} Ys\textsubscript{d}) t * Normal ($\prod_{i\in L3}$ (real (Rel p (X l) t))) *}}}}\\
\mbox{\textup{{\texttt{~~(1 - (1 - Normal ($\prod_{l\in L1}$ (real (Rel p (X l) t)))) *}}}}\\
\mbox{\textup{{\texttt{~~~~~~~(1 - Normal ($\prod_{l\in L2}$ (real (Rel p (X l) t))))))}}}}
\end{theorem}

\begin{figure}[h]
\centering
\includegraphics[scale=1]{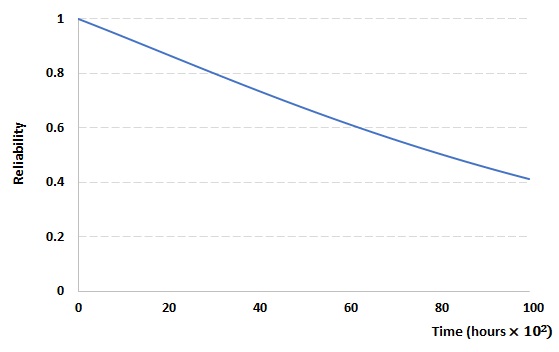}
\caption{Broadcast Reliability of a $128\times 128$ SEN+}
\label{fig:drbd_broad_plot}
\end{figure}

We evaluate the broadcast reliability, in Figure~\ref{fig:drbd_broad_plot}, of a $128\times 128$ SEN+, where each inner series structure of Figure~\ref{fig:broadcast_rel_sen+} has 63 blocks and the series structure on the right hand side of the figure has 64 blocks. We use the same failure rates of  $1\times 10^{-5}$ for each switching element with a dormancy factor of $0.1$.  

\section{Network Reliability Analysis of Shuffle-exchange Networks}
According to \cite{bistouni2014analyzing}, the network reliability of SENs can be defined as the reliability of all connections between sources (inputs) and destinations (outputs). In other words, we are looking at the reliability of the overall network. This is usually modeled using RBDs. In this section, we use both DFT and DRBD models in different scenarios to model the reliability of the network. 

\subsection{DFT Analysis of SEN and SEN+}

\begin{figure}[hbtp]
\centering
\includegraphics[scale=0.7]{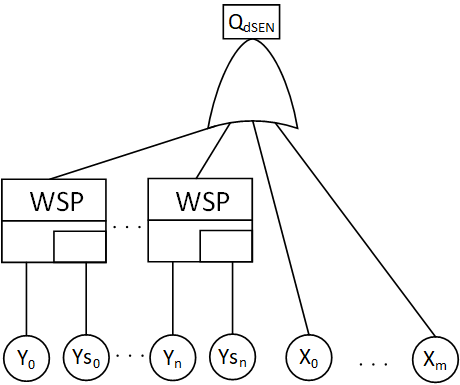}
\caption{DFT of SEN Network with Multiple Spares}
\label{fig:dft_network_sen}
\end{figure}

In the SEN, it is required that all switching elements must work properly in order to maintain a successful behavior of the network. Thus, the system fails with the failure of any of the switching elements. The behavior can be further enhanced by using spares. The DFT of the SEN network can be modeled as in Figure~\ref{fig:dft_sen}. However, to further enhance the system reliability,  the reliability engineer may suggest to use more spares to replace the switching elements. Therefore, we present a generic model, where the number of switching elements that have spares is generic, as shown in Figure~\ref{fig:dft_network_sen}. This model can be also used with both the terminal and broadcast models, when more spares are required.

The top event of the DFT of Figure~\ref{fig:dft_network_sen} can be expressed using the DFT operators as:

\begin{equation}
\begin{split}
\textup{\texttt{Q\textsubscript{dsSEN\_Network} =}}& \textup{\texttt{ n\_OR (MAP ($\lambda$i. if i $\in$ L1 then WSP (Y i) (Ys\textsubscript{a} i) (Ys\textsubscript{d} i)}}\\
&\textup{\texttt{~~~~~~~~~~~~~~~~~else X i) (SET\_TO\_LIST (L1 $\cup$ L2)))}}
\end{split}
\end{equation}

We verify the probability of failure of the top event in a similar way to Theorem~\ref{thm:term_dft_sSEN}, as:

\begin{theorem}
\label{thm:network_dft_sSEN}
\emph{}\\
\textup{{\texttt{$\vdash\forall$ p X Y Ys\textsubscript{a} Ys\textsubscript{d} t L1 L2.}}}\\
\mbox{\textup{\texttt{~DISJOINT L1 L2 $\wedge$ FINITE L1 $\wedge$ L1 $\neq$ \{\} $\wedge$}}}\\
\mbox{\textup{\texttt{~FINITE L2 $\wedge$ L2 $\neq$ \{\} $\wedge$}}}\\
\mbox{\textup{{\texttt{~($\forall$ i. i $\in$ L2 $\Rightarrow$ rv\_gt0\_ninfinity [X i]) $\wedge$ }}}}\\
\mbox{\textup{\texttt{~indep\_sets p}}}\\
\mbox{\textup{\texttt{~~~($\lambda$i.}}}\\
\mbox{\textup{\texttt{~~~~~\{rv\_to\_devent p}}}\\
\mbox{\textup{\texttt{ ~~~~~~~($\lambda$i.  i $\in$ L1 then WSP (Y i) (Ys\textsubscript{a} i) (Ys\textsubscript{d} i) else X i)}}}\\
\mbox{\textup{\texttt{~~~~~~~~t i\})(L1 $\cup$ L2) $\Rightarrow$}}}\\
\mbox{\textup{{\texttt{~(prob p (DFT\_event p Q\textsubscript{dSEN\_Network} t) = }}}}\\
\mbox{\textup{\texttt{~~1-}}}\\
\mbox{\textup{\texttt{~~Normal}}}\\
\mbox{\textup{\texttt{~~~~($\prod_{i\in L1}$}}}\\
\mbox{\textup{\texttt{~~~~~~(real(1- prob p (DFT\_event p (WSP (Y i) (Ys\textsubscript{a} i) (Ys\textsubscript{d} i)) t)))) *}}}\\
\mbox{\textup{\texttt{~~Normal ($\prod_{i\in L2}$ (real(1-F\textsubscript{X\textsubscript{i}}(t)))))}}} 
\end{theorem}

\noindent where \texttt{Y}, \texttt{Ys\textsubscript{a}} and \texttt{Ys\textsubscript{d}} are groups of indexed random variables that represent the main and spare switches. Theorem~\ref{thm:network_dft_sSEN} provides a generic scenario for the SEN, where \texttt{L1} and \texttt{L2} can be instantiated with any number of distinct indices that represent the system switches, with and without spares.

\begin{figure}[hbtp]
\centering
\includegraphics[width=\textwidth]{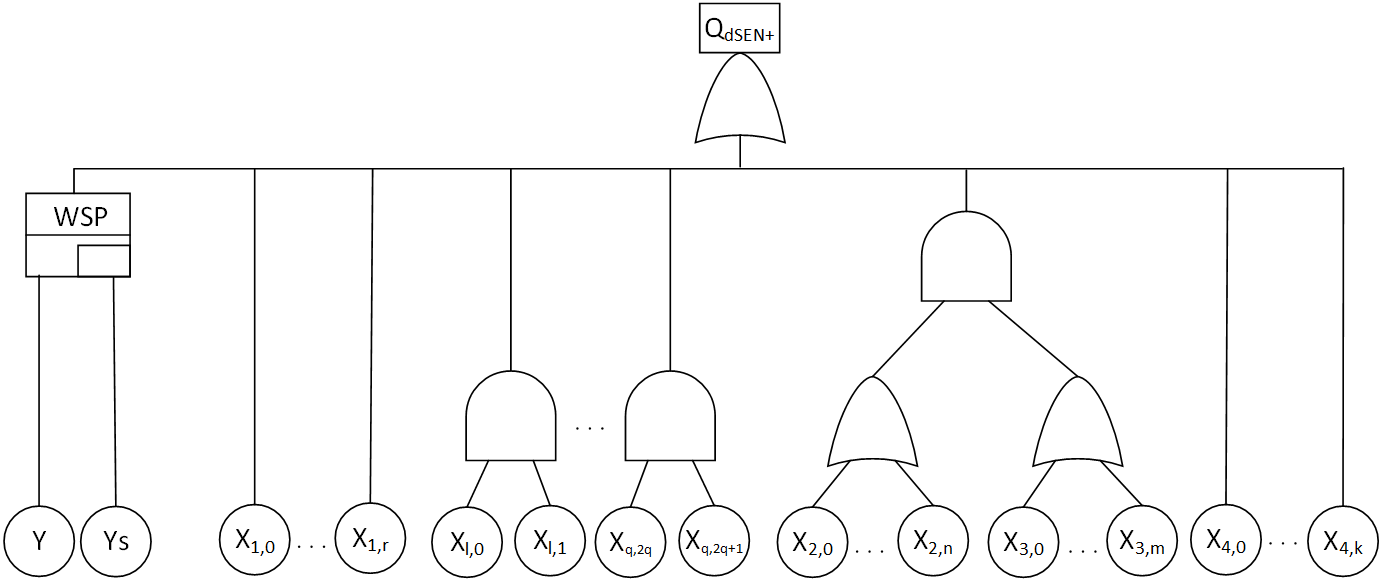}
\caption{DFT of SEN+ Network}
\label{fig:SEN+_network_dft}
\end{figure}

The DFT model of the SEN+ network is shown in Figure~\ref{fig:SEN+_network_dft}. It consists of a spare gate for one of the switches in the input stage. The rest of the input switches ($X_{1,0}$ - $X_{1,r}$) are connected directly to the $n$-OR gate of the top event. Therefore, the failure of any of these switches leads to the failure of the network. The series of ANDs and ANDs of ORs are used to model the two available paths. Finally, all destination switches ($X_{4,0}$ -$X_{4,k}$) are required to function and thus they are all connected to the output OR gate. This DFT is composed of three levels; OR of ANDs of ORs, and thus we can use the theorems of union of intersections of unions to verify its probability of failure if the sets of indices are handled properly. 

We first express the top event using the DFT operators as:

\vspace{-20pt}
\begin{equation}
\begin{split}
&\textup{\texttt{Q\textsubscript{dSEN\_Network} =}}\\
&\textup{\texttt{ n\_OR}}\\
&\textup{\texttt{~~(MAP}}\\
&\textup{\texttt{~~~~($\lambda$i. if i = 0 then WSP Y Ys\textsubscript{a} Ys\textsubscript{d}}}\\
&\textup{\texttt{~~~~~~~~~~else if i = 1 then n\_OR (MAP X (SET\_TO\_LIST L1))}}\\
&\textup{\texttt{~~~~~~~~~~else if i = 3 then (n\_OR (MAP X (SET\_TO\_LIST L2))) $\cdot$}}\\
&\textup{\texttt{~~~~~~~~~~~~~~~~~~~~~~~~~~~~~(n\_OR (MAP X (SET\_TO\_LIST L3)))}}\\
&\textup{\texttt{~~~~~~~~~~else if i = 4 then n\_OR (MAP X (SET\_TO\_LIST L4))}}\\
&\textup{\texttt{~~~~~~~~~~else (X (2 * i)) $\cdot$ (X (2 * i + 1)))}}\\
&\textup{\texttt{~~~~(SET\_TO\_LIST (\{0; 1; 3; 4\} $\cup$ L)))}}
\end{split}
\end{equation}

\noindent where the spare gate is assigned index $0$. The second group of switches has index $1$, while the indices of these switches, $X_{1,0}$ - $X_{1,r}$, are in set \texttt{L1}. They are represented as \texttt{n\_OR (MAP X (SET\_TO\_LIST L1)}. The output of the AND of ORs is assigned index $3$ and is modeled as  \texttt{(n\_OR (MAP X (SET\_TO\_LIST L2))) $\cdot$ (n\_OR (MAP X (SET\_TO\_LIST L3)))}, which is similar to both the terminal and broadcast models. The group of switches, $X_{4,0}$ -$X_{4,k}$, has index $4$ and is represented using \texttt{n\_OR (MAP X (SET\_TO\_LIST L4))}. Thus, we have the indices $\{0; 1; 3; 4\}$ for the outer groups in the DFT. However, the last part of the DFT, which is the series of ANDs in the middle of Figure~\ref{fig:SEN+_network_dft}, has a generic number of AND gates and cannot be assigned a specific index. Therefore, we use set \texttt{L} to get a unique index for the output of each AND gate. We use this unique number to create the indices of the inputs of each AND gate. For example, for an index \texttt{j} in set \texttt{L}, we create two indices for the inputs of the AND gate as \texttt{(2*j)} and \texttt{(2*j+1)}. This is modeled as \texttt{(X (2 * i)) $\cdot$ (X (2 * i + 1)))} and set \texttt{L} is used with the set of indices in the outer level as \texttt{(SET\_TO\_LIST (\{0; 1; 3; 4\} $\cup$ L))}. It is important to highlight that the indices of the individual inputs should be unique. 

We then verify that the \texttt{DFT\_event} of \texttt{Q\textsubscript{dSEN\_Network}} is equal to the union of intersection of union of events as in the following theorem:

\begin{theorem}
\label{thm:SEN+_network_DFT_n_OR}
\emph{}\\
\mbox{\textup{\texttt{$\vdash\forall$ p L1 L2 L3 L4 L X Y Ys\textsubscript{a} Ys\textsubscript{d} t.}}}\\
\mbox{\textup{\texttt{~~FINITE L1 $\wedge$ L1 $\neq$ \{\} $\wedge$ FINITE L2 $\wedge$ L2 $\neq$ \{\} $\wedge$ FINITE L3 $\wedge$}}}\\
\mbox{\textup{\texttt{~~L3 $\neq$ \{\} $\wedge$ FINITE L4 $\wedge$ L4 $\neq$ \{\} $\wedge$ FINITE L $\wedge$}}}\\
\mbox{\textup{\texttt{~~DISJOINT \{0; 1; 3; 4\} L $\wedge$}}}\\
\mbox{\textup{\texttt{~~($\forall$ i. i $\in$ L $\Rightarrow$ DISJOINT \{2 * i; 2 * i + 1\} \{0; 1; 2; 3; 4\}) $\wedge$}}}\\
\mbox{\textup{\texttt{~~disjoint\_family\_on}}}\\
\mbox{\textup{\texttt{~~~~(ind\_set}}}\\
\mbox{\textup{\texttt{~~~~~~[\{0\}; L1; L2; L3; L4; \{2 * i | i $\in$ L\} $\cup$ \{2 * i + 1 | i $\in$ L\}])}}}\\
\mbox{\textup{\texttt{~~~~\{0; 1; 2; 3; 4; 5\} $\Rightarrow$}}}\\
\mbox{\textup{\texttt{~~(DFT\_event p (Q\textsubscript{dSEN\_Network}) t =}}}\\
\mbox{\textup{\texttt{~~~$\bigcup$}}}\\
\mbox{\textup{\texttt{~~~~\{$\bigcap$}}}\\
\mbox{\textup{\texttt{~~~~~~\{$\bigcup$ \{event\_set [(DFT\_event p (WSP Y Ys\textsubscript{a} Ys\textsubscript{d}) t,0)]}}}\\
\mbox{\textup{\texttt{~~~~~~~~~~~~~(rv\_to\_devent p X t) i |}}}\\
\mbox{\textup{\texttt{~~~~~~~~~i $\in$ if a $\in$ \{2 * i | i $\in$ L\} $\cup$ \{2 * i + 1 | i $\in$ L\} then \{a\}}}}\\
\mbox{\textup{\texttt{~~~~~~~~~~~~~~~~else ind\_set [\{0\}; L1; L2; L3; L4] a\} |}}}\\
\mbox{\textup{\texttt{~~~~~a $\in$ if j $\in$ L then \{2 * j; 2 * j + 1\}}}}\\
\mbox{\textup{\texttt{~~~~~~~~~~~else ind\_set [\{0\}; \{1\}; \{\}; \{2; 3\}; \{4\}] j\} |}}}\\
\mbox{\textup{\texttt{~~~j $\in$ \{0; 1; 3; 4\} $\cup$ L\}}}}  

\end{theorem}

\noindent where the conditions are required to ensure that the sets are finite, nonempty and that at each level of the DFT the indices are unique. 
It is clear from the theorem how the hierarchy of the DFT is structured using the sets. For example, ``\texttt{if j $\in$ L then \{2 * j; 2 * j + 1\} else ind\_set [\{0\}; \{1\}; \{\}; \{2; 3\}; \{4\}] j}" determines the indices of the second level of the DFT (the ORs) based on the value of \texttt{j} in the outer level. The first part ``\texttt{if j $\in$ L then \{2 * j; 2 * j + 1\}}" is for the series of ANDs, while ``\texttt{else ind\_set [\{0\}; \{1\}; \{\}; \{2; 3\}; \{4\}] j}" is for the rest of the parts in the second level. Although some of the parts of the DFT have no intermediate OR gates, like the spare, we implicitly assume that there are OR gates with single inputs to maintain the consistency. The indices of the second level indicates the indices of the output of these gates. This can be obvious for the AND of ORs in Figure~\ref{fig:SEN+_network_dft}, where the OR gates have indices $2$ and $3$. We use an empty set (\{\}) in the indices of the second level due to the fact that there is no index $2$ in the outer level, and thus we assigned an empty set in the second level for this index.  

We verify the probability of failure of \texttt{Q\textsubscript{dSEN\_Network}} as:

\begin{theorem}
\label{thm:prob_SEN+_network_dft}
\emph{}\\
\mbox{\textup{\texttt{$\vdash\forall$ p L1 L2 L3 L4 L X Y Ys\textsubscript{a} Ys\textsubscript{d} t.}}}\\
\mbox{\textup{\texttt{~~SEN\_network\_set\_req p L1 L2 L3 L4 L}}}\\
\mbox{\textup{\texttt{~~~~($\lambda$i.}}}\\
\mbox{\textup{\texttt{~~~~~~if i $\in$ \{2 * i | i $\in$ L\} $\cup$ \{2 * i + 1 | i $\in$ L\} then \{i\}}}}\\
\mbox{\textup{\texttt{~~~~~~else ind\_set [\{0\}; L1; L2; L3; L4] i)}}}\\
\mbox{\textup{\texttt{~~~~($\lambda$j.}}}\\
\mbox{\textup{\texttt{~~~~~~if j $\in$ L then \{2 * j; 2 * j + 1\}}}}\\
\mbox{\textup{\texttt{~~~~~~else ind\_set [\{0\}; \{1\}; \{\}; \{2; 3\}; \{4\}] j)}}}\\
\mbox{\textup{\texttt{~~~~(\{0; 1; 3; 4\} $\cup$ L)}}}\\
\mbox{\textup{\texttt{~~~~~(event\_set [(DFT\_event p (WSP Y Ys\textsubscript{a} Ys\textsubscript{d}) t,0)]}}}\\
\mbox{\textup{\texttt{~~~~~~~(rv\_to\_devent p X t)) $\wedge$}}}\\
\mbox{\textup{\texttt{~~($\forall$ i.}}}\\
\mbox{\textup{\texttt{~~~~~i $\in$ L1 $\cup$ L2 $\cup$ L3 $\cup$ L4 $\cup$}}}\\
\mbox{\textup{\texttt{~~~~~~ \{2 * i | i $\in$ L\} $\cup$
               \{2 * i + 1 | i $\in$ L\} $\Rightarrow$}}}\\
\mbox{\textup{\texttt{~~~~~rv\_gt0\_ninfinity [X i]) $\Rightarrow$}}}\\
\mbox{\textup{\texttt{~~(prob p}}}\\
\mbox{\textup{\texttt{~~~(DFT\_event p (Q\textsubscript{dSEN\_Network}) t) =}}}\\
\mbox{\textup{\texttt{~~~1 -}}}\\
\mbox{\textup{\texttt{~~~(1 - prob p (DFT\_event p (WSP Y Ys\textsubscript{a} Ys\textsubscript{d}) t)) *}}}\\
\mbox{\textup{\texttt{~~~Normal ($\prod_{l\in L1}$ (real (1 -  F\textsubscript{X\textsubscript{l}}(t)))) *}}}\\
\mbox{\textup{\texttt{~~~(1 -}}}\\
\mbox{\textup{\texttt{~~~~(1 - Normal ($\prod_{l\in L2}$ (real (1 - F\textsubscript{X\textsubscript{l}}(t))))) *}}}\\
\mbox{\textup{\texttt{~~~~(1 - Normal ($\prod_{l\in L3}$ (real (1 - F\textsubscript{X\textsubscript{l}}(t)))))) *}}}\\
\mbox{\textup{\texttt{~~~~Normal ($\prod_{l\in L4}$ (real (1 - F\textsubscript{X\textsubscript{l}}(t)))) *}}}\\
\mbox{\textup{\texttt{~~~~Normal
             ($\prod_{j\in L}$                (1 - real
                       (F\textsubscript{X\textsubscript{2*j}}(t) *
                        F\textsubscript{X\textsubscript{2*j+1}}(t)))))}}}

\end{theorem}

\noindent where \texttt{SEN\_network\_set\_req} ensures all the required conditions for the sets to be finite, nonempty and distinct. It also ensures the independence of the input events. It accepts all the sets of the indices of the three levels. The second condition (\texttt{rv\_gt0\_ninfinity [X i]}) ascertains that each element in the group of random variables of \texttt{X} that have their indices in \texttt{L1 $\cup$ L2 $\cup$ L3 $\cup$ L4 $\cup$ \{2 * i | i $\in$ L\} $\cup$ \{2 * i + 1 | i $\in$ L\} } are greater than or equal to $0$ but not equal $+\infty$. This condition is required to be able to use the CDF of the random variables.

\begin{figure}[hbtp]
\centering
\includegraphics[width=\textwidth]{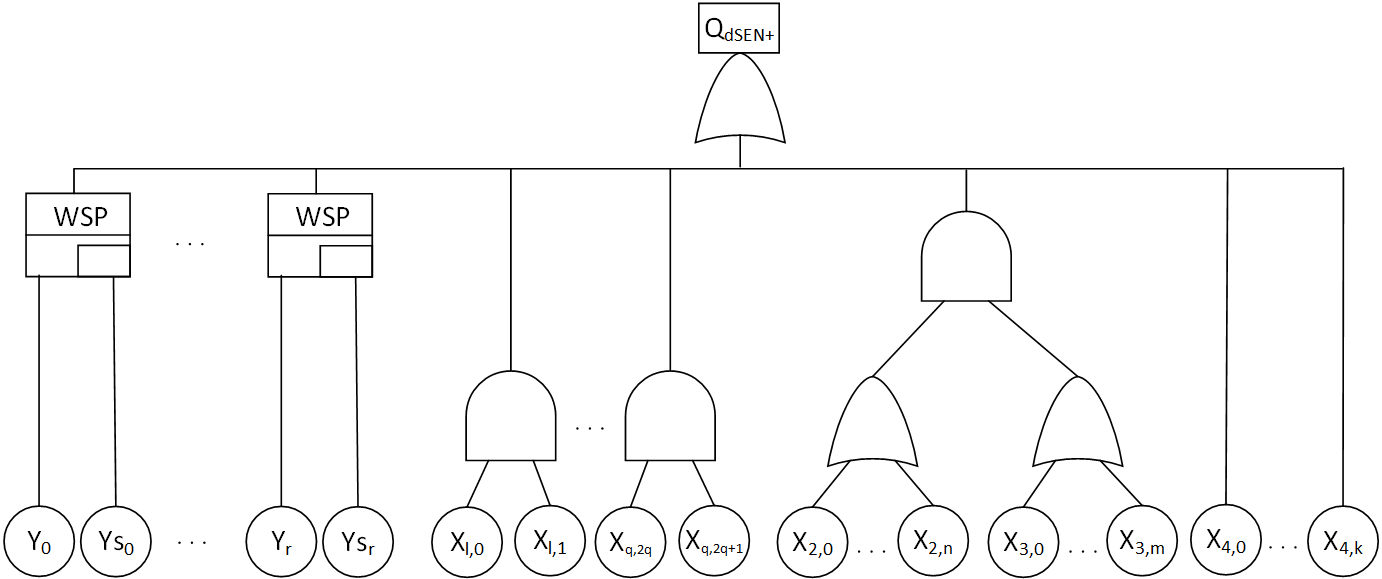}
\caption{DFT of SEN+ with Multiple Spares}
\label{fig:SEN+_network_dft_spares}
\end{figure}

In a similar manner to the SEN network, we provide a generic model where any number of spares can be used for the input switches. The modified DFT is shown in Figure~\ref{fig:SEN+_network_dft_spares}. We express the top event using the DFT operators as:

\vspace{-20pt}
\begin{equation}
\begin{split}
&\textup{\texttt{Q\textsubscript{dSEN\_Network2} =}}\\
&\textup{\texttt{ n\_OR}}\\
&\textup{\texttt{~~(MAP}}\\
&\textup{\texttt{~~~~($\lambda$i. if i = 0 then WSP (Y 0) (Ys\textsubscript{a} 0) (Ys\textsubscript{d} 0)}}\\
&\textup{\texttt{~~~~~~~~~~~else if i = 1 then}}\\
&\textup{\texttt{~~~~~~~~~~~~(n\_OR (MAP X (SET\_TO\_LIST L1))) }}\\
&\textup{\texttt{~~~~~~~~~~else if i = 3 then (n\_OR (MAP X (SET\_TO\_LIST L2))) $\cdot$}}\\
&\textup{\texttt{~~~~~~~~~~~~~~~~~~~~~~~~~~~~~(n\_OR (MAP X (SET\_TO\_LIST L3)))}}\\
&\textup{\texttt{~~~~~~~~~~else if i = 4 then n\_OR (MAP X (SET\_TO\_LIST L4))}}\\
&\textup{\texttt{~~~~~~~~~~else (X (2 * i)) $\cdot$ (X (2 * i + 1)))}}\\
&\textup{\texttt{~~~~(SET\_TO\_LIST (\{0; 1; 3; 4\} UNION L)))}}
\end{split}
\end{equation}

\noindent where \texttt{Y}, \texttt{Ys\textsubscript{a}} and \texttt{Ys\textsubscript{d}} are indexed random variables that represent the main and spare parts for each spare gate. We choose to use the same hierarchy of Figure~\ref{fig:SEN+_network_dft}, where we assign index $0$ for the first spare and the rest of the spares have their indices in set \texttt{L1}. In addition, the model of these additional spares is embedded within \texttt{X} as will be explained shortly. 

We verify the probability of failure of the top event as:

\begin{theorem}
\label{thm:prob_SEN+_network_dft_spares}
\emph{}\\
\mbox{\textup{\texttt{$\vdash\forall$ p L1 L2 L3 L4 L X Y Ys\textsubscript{a} Ys\textsubscript{d} t.}}}\\
\mbox{\textup{\texttt{~~SEN\_network\_set\_req p L1 L2 L3 L4 L}}}\\
\mbox{\textup{\texttt{~~~~($\lambda$i.}}}\\
\mbox{\textup{\texttt{~~~~~~if i $\in$ \{2 * i | i $\in$ L\} $\cup$ \{2 * i + 1 | i $\in$ L\} then \{i\}}}}\\
\mbox{\textup{\texttt{~~~~~~else ind\_set [\{0\}; L1; L2; L3; L4] i)}}}\\
\mbox{\textup{\texttt{~~~~($\lambda$j.}}}\\
\mbox{\textup{\texttt{~~~~~~if j $\in$ L then \{2 * j; 2 * j + 1\}}}}\\
\mbox{\textup{\texttt{~~~~~~else ind\_set [\{0\}; \{1\}; \{\}; \{2; 3\}; \{4\}] j)}}}\\
\mbox{\textup{\texttt{~~~~(\{0; 1; 3; 4\} $\cup$ L)}}}\\
\mbox{\textup{\texttt{~~~~($\lambda$i.}}}\\
\mbox{\textup{\texttt{~~~~~~event\_set [(DFT\_event p (WSP (Y 0) (Ys\textsubscript{a} 0) (Ys\textsubscript{d} 0)) t,0)]}}}\\
\mbox{\textup{\texttt{~~~~~~~(rv\_to\_devent p X t) i) $\wedge$}}}\\
\mbox{\textup{\texttt{~~($\forall$ i.}}}\\
\mbox{\textup{\texttt{~~~i $\in$ L1 $\cup$ L2 $\cup$ L3 $\cup$ L4 $\cup$ \{2 * i | i $\in$ L\} $\cup$
               \{2 * i + 1 | i $\in$ L\} $\Rightarrow$}}}\\
\mbox{\textup{\texttt{~~~rv\_gt0\_ninfinity [X i]) $\wedge$ }}}\\
\mbox{\textup{\texttt{~~($\forall$ i. i $\in$ L1 $\Rightarrow$ (X i = WSP (Y i) (Ys\textsubscript{a} i) (Ys\textsubscript{d} i))$\Rightarrow$}}}\\
\mbox{\textup{\texttt{~~(prob p}}}\\
\mbox{\textup{\texttt{~~~(DFT\_event p (Q\textsubscript{dSEN\_Network2}) t) =}}}\\
\mbox{\textup{\texttt{~~~1 -}}}\\
\mbox{\textup{\texttt{~~~Normal}}}\\
\mbox{\textup{\texttt{~~~~($\prod_{l\in (\{0\}\cup L1}$}}}\\
\mbox{\textup{\texttt{~~~~~(real (1 - prob p (DFT\_event p (WSP (Y l) (Ys\textsubscript{a} l) (Ys\textsubscript{d} l)) t))) *}}}\\
\mbox{\textup{\texttt{~~~(1 -}}}\\
\mbox{\textup{\texttt{~~~~(1 - Normal ($\prod_{l\in L2}$ (real (1 - F\textsubscript{X\textsubscript{l}}(t))))) *}}}\\
\mbox{\textup{\texttt{~~~~(1 - Normal ($\prod_{l\in L3}$ (real (1 - F\textsubscript{X\textsubscript{l}}(t)))))) *}}}\\
\mbox{\textup{\texttt{~~~~Normal ($\prod_{l\in L4}$ (real (1 - F\textsubscript{X\textsubscript{l}}(t)))) *}}}\\
\mbox{\textup{\texttt{~~~~Normal
             ($\prod_{j\in L}$                (1 - real
                       (F\textsubscript{X\textsubscript{2*j}}(t) *
                        F\textsubscript{X\textsubscript{2*j+1}}(t)))))}}}
\end{theorem}

\noindent where the conditions are similar to Theorem~\ref{thm:prob_SEN+_network_dft}. However, we add the condition that \texttt{($\forall$ i. i $\in$ L1 $\Rightarrow$ (X i = WSP (Y i) (Ys\textsubscript{a} i) (Ys\textsubscript{d} i))}, which adds the additional spare gates. This way, we can use Theorem~\ref{thm:prob_SEN+_network_dft} to verify Theorem~\ref{thm:prob_SEN+_network_dft_spares}. Set \texttt{\{0\} $\cup$ L1} is used to provide the indices of the spares, including the first one with index $0$.

We evaluate the probability of failure of the network DFT, shown in Figure~\ref{fig:SEN+_network_dft_spares}, for a $128\times 128$ SEN+. The DFT of this SEN has $32$ AND gates in the first level. Each OR gate in the first level has $160$ inputs. Furthermore, we assume that all the $64$ input switches have spares. Figure~\ref{fig:dft_network_plot} shows the evaluated result of the probability of failure, where the failure rates of each switching element is $1\times 10^{-5}$ with a dormancy factor of $0.1$.  

\begin{figure}[hbtp]
\centering
\includegraphics[scale=1]{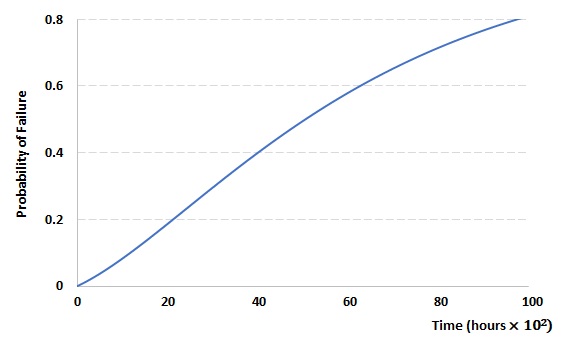}
\caption{The Probability of Failure of the Network of a $128\times 128$ SEN+}
\label{fig:dft_network_plot}
\end{figure}

\subsection{DRBD Analysis of SEN and SEN+}

\begin{figure}[hbtp]
\centering
\includegraphics[scale=0.8]{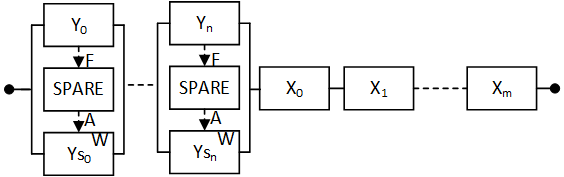}
\caption{DRBD of SEN Network}
\label{fig:SEN_drbd_network}
\end{figure}

Similar to the DFT models, we start first with the network reliability model of the SEN. Since it is a single path, it can be modeled using the series DRBD of Figure~\ref{fig:term_sen_DRBD}. Thus, we can use Theorem~\ref{thm:term_Rel_sSEN} to provide a generic expression for its reliability. We provide a generic model in Figure~\ref{fig:SEN_drbd_network}, where additional spares are used. This provides a general case where we can choose how many switches can be replaced with spares. 

We express the structure function of this DRBD using our DRBD operators as:

\begin{equation}
\begin{split}
\textup{\texttt{Q\textsubscript{sSEN\_Network} =}}& \textup{\texttt{ nR\_AND ($\lambda$i. if i $\in$ L1 then R\_WSP (Y i) (Ys\textsubscript{a} i) (Ys\textsubscript{d} i)}}\\
&\textup{\texttt{~~~~~~~~~~~~~~else X i) (L1 $\cup$ L2)}}
\end{split}
\end{equation}

\noindent where \texttt{L1} and \texttt{L2} provide the indices of the blocks in the series structure for the spare constructs and the remaining blocks, respectively. 

Similar to the proof steps of Theorem~\ref{thm:Rel_SEN}, we verify the reliability of the SEN network as:

\begin{theorem}
\label{thm:network_drbd_sSEN}
\emph{}\\
\textup{{\texttt{$\vdash\forall$ p X Y Ys\textsubscript{a} Ys\textsubscript{d} t L1 L2.}}}\\
\mbox{\textup{\texttt{~DISJOINT L1 L2 $\wedge$ FINITE L1 $\wedge$ L1 $\neq$ \{\} $\wedge$}}}\\
\mbox{\textup{\texttt{~FINITE L2 $\wedge$ L2 $\neq$ \{\} $\wedge$}}}\\
\mbox{\textup{\texttt{~indep\_sets p}}}\\
\mbox{\textup{\texttt{~~~($\lambda$i. \{if i $\in$ L1 then DRBD\_event p (R\_WSP (Y i) (Ys\textsubscript{a} i) (Ys\textsubscript{d} i)) t }}}\\
\mbox{\textup{\texttt{~~~~~~~~~~else (rv\_to\_event p X t) i\}) (L1 $\cup$ L2)$\Rightarrow$}}}\\
\mbox{\textup{{\texttt{~(prob p (DRBD\_event p Q\textsubscript{SEN\_Network} t) = }}}}\\
\mbox{\textup{\texttt{~~Normal}}}\\
\mbox{\textup{\texttt{~~~~($\prod_{i\in L1}$}}}\\
\mbox{\textup{\texttt{~~~~~~(real (Rel p (R\_WSP (Y i) (Ys\textsubscript{a} i) (Ys\textsubscript{d} i)) t))) *}}}\\
\mbox{\textup{\texttt{~~Normal ($\prod_{i\in L2}$ (real (Rel p (X i) t))))}}} 
\end{theorem}

\begin{figure}[!t]
\centering
\includegraphics[width=\textwidth]{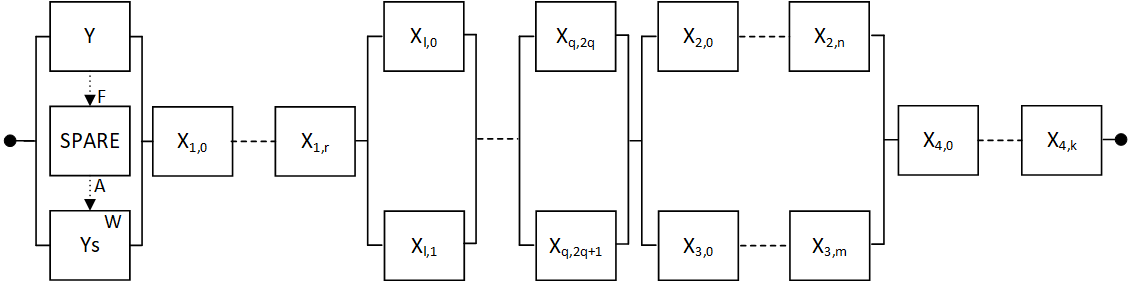}
\caption{DRBD of SEN+ Newtork }
\label{fig:SEN+_network_drbd}
\end{figure}

The DRBD of the SEN+ network is modeled in Figure~\ref{fig:SEN+_network_drbd}, where only one of the switches of the input stage can be replaced by a spare.
This DRBD is composed of a series-parallel-series structure. The indices of each level can be treated in a similar manner to the DFT. 

We express the structure function using the operators with the same sets of indices of the DFT as:

\vspace{-20pt}
\begin{equation}
\begin{split}
\textup{\texttt{Q\textsubscript{SEN\_Network} =}}& \textup{\texttt{ nR\_AND }}\\
&\textup{\texttt{~~~($\lambda$i.}}\\
&\textup{\texttt{~~~~~if i = 0 then R\_WSP Y Ys\textsubscript{a} Ys\textsubscript{d}}}\\
&\textup{\texttt{~~~~~else if i = 1 then nR\_AND X L1}}\\
&\textup{\texttt{~~~~~else if i = 3 then (nR\_AND X L2) + (nR\_AND X L3)}}\\
&\textup{\texttt{~~~~~else if i = 4 then nR\_AND X L4}}\\
&\textup{\texttt{~~~~~else (X (2 * i)) + (X (2 * i + 1)))}}\\
&\textup{\texttt{~~~(\{0; 1; 3; 4\} $\cup$ L))}}
\end{split}
\end{equation}

Then, we verify that the \texttt{DRBD\_event} of this structure can be expressed as a series-parallel-series structure as:

\begin{theorem}
\emph{}\\
\label{thm:SEN+_drbd_network_nR_AND}
\mbox{\textup{\texttt{$\vdash\forall$ p L1 L2 L3 L4 L X Y Ys\textsubscript{a} Ys\textsubscript{d} t.}}}\\
\mbox{\textup{\texttt{~~FINITE L1 $\wedge$ L1 $\neq$ \{\} $\wedge$ FINITE L2 $\wedge$ L2 $\neq$ \{\} $\wedge$ FINITE L3 $\wedge$}}}\\
\mbox{\textup{\texttt{~~L3 $\neq$ \{\} $\wedge$ FINITE L4 $\wedge$ L4 $\neq$ \{\} $\wedge$ FINITE L $\wedge$}}}\\
\mbox{\textup{\texttt{~~DISJOINT \{0; 1; 3; 4\} L $\wedge$}}}\\
\mbox{\textup{\texttt{~~($\forall$ i. i $\in$ L $\Rightarrow$ DISJOINT \{2 * i; 2 * i + 1\} \{0; 1; 2; 3; 4\}) $\wedge$}}}\\
\mbox{\textup{\texttt{~~disjoint\_family\_on}}}\\
\mbox{\textup{\texttt{~~~(ind\_set}}}\\
\mbox{\textup{\texttt{~~~~~[\{0\}; L1; L2; L3; L4;}}}\\
\mbox{\textup{\texttt{~~~~~~\{2 * i | i $\in$ L\} $\cup$ \{2 * i + 1 | i $\in$ L\}])}}}\\
\mbox{\textup{\texttt{~~~~~\{0; 1; 2; 3; 4; 5\} $\Rightarrow$}}}\\
\mbox{\textup{\texttt{~~(DRBD\_event p}}}\\
\mbox{\textup{\texttt{~~~(Q\textsubscript{SEN\_Network}) t =}}}\\
\mbox{\textup{\texttt{~~~DRBD\_series}}}\\
\mbox{\textup{\texttt{~~~~($\lambda$j.}}}\\
\mbox{\textup{\texttt{~~~~~~DRBD\_parallel}}}\\
\mbox{\textup{\texttt{~~~~~~~~($\lambda$a.}}}\\
\mbox{\textup{\texttt{~~~~~~~~~~DRBD\_series}}}\\
\mbox{\textup{\texttt{~~~~~~~~~~~~($\lambda$i.}}}\\
\mbox{\textup{\texttt{~~~~~~~~~~~~~~~event\_set}}}\\
\mbox{\textup{\texttt{~~~~~~~~~~~~~~~~~[(DRBD\_event p (R\_WSP Y Ys\textsubscript{a} Ys\textsubscript{d} t,0)]}}}\\
\mbox{\textup{\texttt{~~~~~~~~~~~~~~~~~(rv\_to\_event p X t) i)}}}\\
\mbox{\textup{\texttt{~~~~~~~~~~~~(($\lambda$i.}}}\\
\mbox{\textup{\texttt{~~~~~~~~~~~~~~~if i $\in$ \{2 * i | i $\in$ L\} $\cup$ \{2 * i + 1 | i $\in$ L\} then \{i\}}}}\\
\mbox{\textup{\texttt{~~~~~~~~~~~~~~~else ind\_set [\{0\}; L1; L2; L3; L4] i) a))}}}\\
\mbox{\textup{\texttt{~~~~~~~(($\lambda$j.}}}\\
\mbox{\textup{\texttt{~~~~~~~~~~~~if j $\in$ L then \{2 * j; 2 * j + 1\}}}}\\
\mbox{\textup{\texttt{~~~~~~~~~~~else ind\_set [\{0\}; \{1\}; \{\}; \{2; 3\}; \{4\}] j) j))}}}\\
\mbox{\textup{\texttt{~~~~~~~(\{0; 1; 3; 4\} $\cup$ L))}}}
\end{theorem}

Finally, we verify the reliability of the DRBD as:

\begin{theorem}
\emph{}\\
\label{thm:prob_drbd_SEN+_network}
\mbox{\textup{\texttt{$\vdash\forall$ p L1 L2 L3 L4 L X Y Ys\textsubscript{a} Ys\textsubscript{d} t.}}}\\
\mbox{\textup{\texttt{~~SEN\_network\_set\_req p L1 L2 L3 L4 L}}}\\
\mbox{\textup{\texttt{~~~~($\lambda$i.}}}\\
\mbox{\textup{\texttt{~~~~~~if i $\in$ \{2 * i | i $\in$ L\} $\cup$ \{2 * i + 1 | i $\in$ L\} then \{i\}}}}\\
\mbox{\textup{\texttt{~~~~~~else ind\_set [\{0\}; L1; L2; L3; L4] i)}}}\\
\mbox{\textup{\texttt{~~~~($\lambda$j.}}}\\
\mbox{\textup{\texttt{~~~~~~if j $\in$ L then \{2 * j; 2 * j + 1\}}}}\\
\mbox{\textup{\texttt{~~~~~~else ind\_set [\{0\}; \{1\}; \{\}; \{2; 3\}; \{4\}] j)}}}\\
\mbox{\textup{\texttt{~~~~(\{0; 1; 3; 4\} $\cup$ L)}}}\\
\mbox{\textup{\texttt{~~~~(event\_set [(DRBD\_event p (R\_WSP Y Ys\textsubscript{a} Ys\textsubscript{d}) t,0)]}}}\\
\mbox{\textup{\texttt{~~~~~~~(rv\_to\_event p X t)) $\Rightarrow$}}}\\
\mbox{\textup{\texttt{~~(prob p}}}\\
\mbox{\textup{\texttt{~~~~(DRBD\_event p (Q\textsubscript{SEN\_Network}) t) =}}}\\
\mbox{\textup{\texttt{~~~Rel p (R\_WSP Y Ys\textsubscript{a} Ys\textsubscript{d}) t *}}}\\
\mbox{\textup{\texttt{~~~Normal ($\prod_{l\in L1}$ (real (Rel p (X l) t))) *}}}\\
\mbox{\textup{\texttt{~~~(1 -}}}\\
\mbox{\textup{\texttt{~~~~(1 - Normal ($\prod_{l\in L2}$ (real (Rel p (X l) t)))) *}}}\\
\mbox{\textup{\texttt{~~~~(1 - Normal ($\prod_{l\in L3}$ (real (Rel p (X l) t))))) *}}}\\
\mbox{\textup{\texttt{~~~Normal ($\prod_{l\in L4}$(real (Rel p (X l) t))) *}}}\\
\mbox{\textup{\texttt{~~~Normal}}}\\
\mbox{\textup{\texttt{~~~~~($\prod_{j\in L}$}}}\\
\mbox{\textup{\texttt{~~~~~~~(1 -}}}\\
\mbox{\textup{\texttt{~~~~~~~~real}}}\\
\mbox{\textup{\texttt{~~~~~~~~~~((1 - Rel p (X (2 * j)) t) *}}}\\
\mbox{\textup{\texttt{~~~~~~~~~~~(1 - Rel p (X (2 * j + 1)) t)))))}}}
\end{theorem}

It is worth mentioning that the conditions of the sets are similar to Theorem~\ref{thm:prob_SEN+_network_dft} of the DFT.

\begin{figure}[hbtp]
\centering
\vspace{20pt}
\includegraphics[width=\textwidth]
{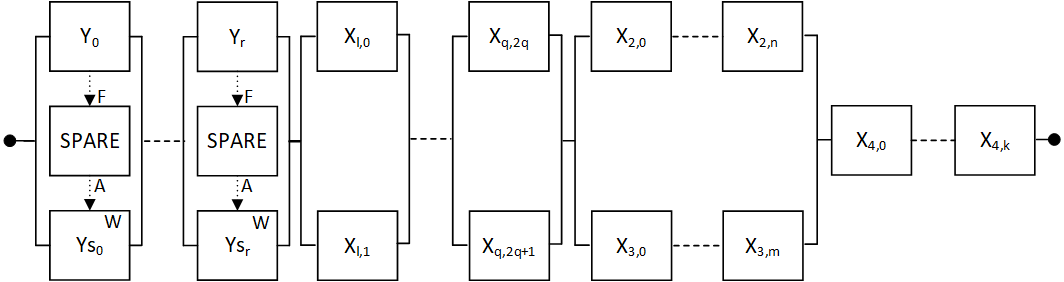}
\caption{DRBD of SEN+ Network with Multiple Spares}
\label{fig:SEN+_network_drbd_spares}
\end{figure}

Finally, we provide a generic model to have any number of spares that can replace the input switches as shown in Figure~\ref{fig:SEN+_network_drbd_spares}. We choose to use the same indices of Figure~\ref{fig:SEN+_network_drbd} in order to reutilize the verified theorems. 

We express the structure of the DRBD of Figure~\ref{fig:SEN+_network_drbd_spares} as:

\begin{equation}
\begin{split}
\textup{\texttt{Q\textsubscript{SEN\_Network2} =}}& \textup{\texttt{ nR\_AND }}\\
&\textup{\texttt{~~~($\lambda$i.}}\\
&\textup{\texttt{~~~~~if i = 0 then R\_WSP (Y 0) (Ys\textsubscript{a} 0) (Ys\textsubscript{d} 0)}}\\
&\textup{\texttt{~~~~~else if i = 1 then nR\_AND X L1}}\\
&\textup{\texttt{~~~~~else if i = 3 then (nR\_AND X L2) + (nR\_AND X L3)}}\\
&\textup{\texttt{~~~~~else if i = 4 then nR\_AND X L4}}\\
&\textup{\texttt{~~~~~else (X (2 * i)) + (X (2 * i + 1)))}}\\
&\textup{\texttt{~~~(\{0; 1; 3; 4\} $\cup$ L))}}
\end{split}
\end{equation}

\noindent where \texttt{(Y 0), (Ys\textsubscript{a} 0)} and \texttt{(Ys\textsubscript{d} 0)} are indexed groups of random variables that represent the main parts and their spares. 

Finally, we use Theorem~\ref{thm:prob_drbd_SEN+_network} to verify the reliability of this DRBD as:

\begin{theorem}
\emph{}\\
\label{thm:prob_drbd_SEN+_network2}
\mbox{\textup{\texttt{$\vdash\forall$ p L1 L2 L3 L4 L X Y Ys\textsubscript{a} Ys\textsubscript{d} t.}}}\\
\mbox{\textup{\texttt{~~SEN\_network\_set\_req p L1 L2 L3 L4 L}}}\\
\mbox{\textup{\texttt{~~~~($\lambda$i.}}}\\
\mbox{\textup{\texttt{~~~~~~if i $\in$ \{2 * i | i $\in$ L\} $\cup$ \{2 * i + 1 | i $\in$ L\} then \{i\}}}}\\
\mbox{\textup{\texttt{~~~~~~else ind\_set [\{0\}; L1; L2; L3; L4] i)}}}\\
\mbox{\textup{\texttt{~~~~($\lambda$j.}}}\\
\mbox{\textup{\texttt{~~~~~~if j $\in$ L then \{2 * j; 2 * j + 1\}}}}\\
\mbox{\textup{\texttt{~~~~~~else ind\_set [\{0\}; \{1\}; \{\}; \{2; 3\}; \{4\}] j)}}}\\
\mbox{\textup{\texttt{~~~~(\{0; 1; 3; 4\} $\cup$ L)}}}\\
\mbox{\textup{\texttt{~~~~(event\_set [(DRBD\_event p (R\_WSP (Y 0) (Ys\textsubscript{a} 0) (Ys\textsubscript{d} 0)) t,0)]}}}\\
\mbox{\textup{\texttt{~~~~~~~(rv\_to\_event p X t)) $\wedge$}}}\\
\mbox{\textup{\texttt{($\forall$ i. i $\in$ L1 $\Rightarrow$ (X i = R\_WSP (Y i) (Ys\textsubscript{a} i) (Ys\textsubscript{d} i))) $\Rightarrow$}}}\\
\mbox{\textup{\texttt{~~(prob p}}}\\
\mbox{\textup{\texttt{~~~~(DRBD\_event p (Q\textsubscript{SEN\_Network2}) t) =}}}\\
\mbox{\textup{\texttt{~~~Normal}}}\\
\mbox{\textup{\texttt{~~~($\prod_{l\in (\{0\}\cup L1}$}}}\\
\mbox{\textup{\texttt{~~~~~~(real (Rel p (R\_WSP (Y l) (Ys\textsubscript{a} l) (Ys\textsubscript{d} l)) t))) *}}}
\mbox{\textup{\texttt{~~~(1 -}}}\\
\mbox{\textup{\texttt{~~~~(1 - Normal ($\prod_{l\in L2}$ (real (Rel p (X l) t)))) *}}}\\
\mbox{\textup{\texttt{~~~~(1 - Normal ($\prod_{l\in L3}$ (real (Rel p (X l) t))))) *}}}\\
\mbox{\textup{\texttt{~~~Normal ($\prod_{l\in L4}$(real (Rel p (X l) t))) *}}}\\
\mbox{\textup{\texttt{~~~Normal}}}\\
\mbox{\textup{\texttt{~~~~~($\prod_{j\in L}$}}}\\
\mbox{\textup{\texttt{~~~~~~~(1 -}}}\\
\mbox{\textup{\texttt{~~~~~~~~real}}}\\
\mbox{\textup{\texttt{~~~~~~~~~~((1 - Rel p (X (2 * j)) t) *}}}\\
\mbox{\textup{\texttt{~~~~~~~~~~~(1 - Rel p (X (2 * j + 1)) t)))))}}}
\end{theorem}

We evaluate the network reliability of a $128\times 128$ as shown in Figure~\ref{fig:SEN+_network_drbd_spares}. In Figure~\ref{fig:SEN+_network_drbd_spares}, there are $32$ parallel structures that are connected in series. The DRBD has $64$ spare constructs, while there are $160$ blocks in the inner series structures. Finally, the series structure on the right hand side of Figure~\ref{fig:SEN+_network_drbd_spares} has $64$ blocks. We assume that the failure rates of each switching element is $1\times 10^{-5}$ with a dormancy factor of $0.1$.  

\begin{figure}[hbtp]
\centering
\includegraphics[scale=1]{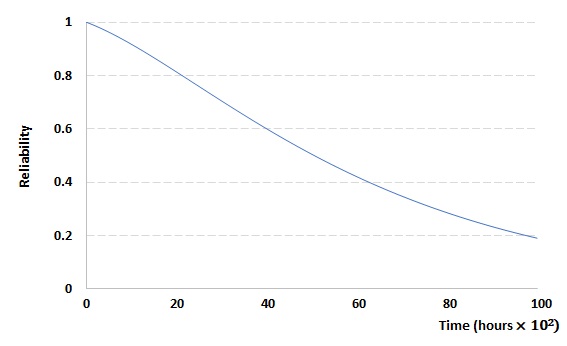}
\caption{The Network Reliability of a $128\times 128$ SEN+}
\label{fig:dft_network_plot}
\end{figure}

\section{Equivalence of SEN DFT and DRBD Models}

In~\cite{elderhalli2019integrating}, we proposed a methodology for where a DFT model can be formally analyzed using the DRBD algebra and vice versa. To illustrate the utilization of the proposed  methodology, we formally verify the equivalence of the DRBD and the complement of the DFT events for both terminal and broadcast reliability of SEN and SEN+. The equivalence of the network models can be conducted in a similar manner.  Proving this equivalence allows verifying the probability of one model and directly use the equivalence proof to provide the probability of the other model. 

We verify the equivalence of the DRBD and DFT models of the terminal reliability of both SEN and SEN+ as:

\begin{theorem}
\textit{Terminal/Broadcast SEN}\\
\label{thm:sen_eq}
\textup{{\texttt{$\vdash\forall$ p X Y Ys\textsubscript{a} Ys\textsubscript{d} t L.}}}\\
\mbox{\textup{{\texttt{~FINITE L $\wedge$ ($\forall$ s. ALL\_DISTINCT [Y s; Ys\textsubscript{a} s; Ys\textsubscript{d} s]) $\Rightarrow$}}}}\\
\mbox{\textup{{\texttt{~(DRBD\_event p}}}}\\
\mbox{\textup{\texttt{~~~~~~(nR\_AND}}}\\
\mbox{\textup{\texttt{~~~~~~~~($\lambda$i.}}}\\
\mbox{\textup{\texttt{~~~~~~~~~~~if i = 0 then R\_WSP Y Ys\textsubscript{a} Ys\textsubscript{d}}}}\\
\mbox{\textup{\texttt{~~~~~~~~~~~else X i) \{0\} $\cup$ L) t =}}}\\
\mbox{\textup{\texttt{~~p\_space p DIFF}}}\\
\mbox{\textup{\texttt{~~DFT\_event p }}}\\
\mbox{\textup{\texttt{~~~~(n\_OR}}}\\
\mbox{\textup{\texttt{~~~~~~(MAP}}}\\
\mbox{\textup{\texttt{~~~~~~~~($\lambda$i.}}}\\
\mbox{\textup{\texttt{~~~~~~~~~~~~if i = 0 then WSP Y Ys\textsubscript{a} Ys\textsubscript{d}}}}\\
\mbox{\textup{\texttt{~~~~~~~~~~~~else X i)}}}\\
\mbox{\textup{\texttt{~~~~~~(SET\_TO\_LIST (\{0\} $\cup$ L)))) t)}}}\\
\end{theorem}

\begin{theorem}
\textit{Terminal SEN+}\\
\textup{{\texttt{$\vdash\forall$ p X Y Ys\textsubscript{a} Ys\textsubscript{d} Z Zs\textsubscript{a} Zs\textsubscript{d} t L1 L2.}}}\\
\mbox{\textup{{\texttt{~FINITE L1 $\wedge$ FINITE L2 $\wedge$ }}}}\\
\mbox{\textup{\texttt{~($\forall$ s. ALL\_DISTINCT [Y s; Ys\textsubscript{a} s; Ys\textsubscript{d} s; Z s; Zs\textsubscript{a} s; Zs\textsubscript{d} s]) $\Rightarrow$}}}\\
\mbox{\textup{{\texttt{~(DRBD\_event p}}}}\\
\mbox{\textup{\texttt{~~~~(nR\_AND}}}\\
\mbox{\textup{\texttt{~~~~~~($\lambda$i.}}}\\
\mbox{\textup{\texttt{~~~~~~~~~~if i = 0 then R\_WSP Y Ys\textsubscript{a} Ys\textsubscript{d}}}}\\
\mbox{\textup{\texttt{~~~~~~~~~~else if i = 1 then \big((nR\_AND X L1) + (nR\_AND X L2)\big)}}}\\
\mbox{\textup{\texttt{~~~~~~~~~~else R\_WSP Z Zs\textsubscript{a} Zs\textsubscript{d}) \{0; 1; 2\}) t = }}}\\
\mbox{\textup{\texttt{~~p\_space p DIFF}}}\\
\mbox{\textup{\texttt{~~DFT\_event p}}}\\
\mbox{\textup{\texttt{~~~~~(n\_OR}}}\\
\mbox{\textup{\texttt{~~~~~~~(MAP}}}\\
\mbox{\textup{\texttt{~~~~~~~~~($\lambda$i.}}}\\
\mbox{\textup{\texttt{~~~~~~~~~~~~if i = 0 then WSP Y Ys\textsubscript{a} Ys\textsubscript{d}}}}\\
\mbox{\textup{\texttt{~~~~~~~~~~~~else if i = 1 then}}}\\
\mbox{\textup{\texttt{~~~~~~~~~~~~~~~~~\big((n\_OR (MAP X (SET\_TO\_LIST L1))) $\cdot$}}}\\
\mbox{\textup{\texttt{~~~~~~~~~~~~~~~~~~~(n\_OR (MAP X (SET\_TO\_LIST L2)))\big)}}}\\
\mbox{\textup{\texttt{~~~~~~~~~~~~else WSP Z Zs\textsubscript{a} Zs\textsubscript{d}) (SET\_TO\_LIST \{0; 1; 2\}))) t)}}}\\

\end{theorem}

In a similar manner, we verify the equivalence of the DRBD and DFT models of the SEN+ broadcast reliability as:\\

\emph{}\\

\begin{theorem}
\textit{Broadcast SEN+}\\
\textup{{\texttt{$\vdash\forall$ p X Y Ys\textsubscript{a} Ys\textsubscript{d} t L1 L2 s.}}}\\
\mbox{\textup{\texttt{~FINITE L1 $\wedge$ FINITE L2 $\wedge$ FINITE L3 $\wedge$ }}}\\
\mbox{\textup{\texttt{~($\forall$ s. ALL\_DISTINCT [Y s; Ys\textsubscript{a} s; Ys\textsubscript{d} s]) $\Rightarrow$}}}\\
\mbox{\textup{\texttt{~(DRBD\_event p}}}\\
\mbox{\textup{\texttt{~~~(nR\_AND }}}\\
\mbox{\textup{\texttt{~~~~~($\lambda$i.}}}\\
\mbox{\textup{\texttt{~~~~~~~~~if i = 0 then R\_WSP Y Ys\textsubscript{a} Ys\textsubscript{d}}}}\\
\mbox{\textup{\texttt{~~~~~~~~~else if i = 1 then \big((nR\_AND X L1) $+$ (nR\_AND X L2)\big)}}}\\
\mbox{\textup{\texttt{~~~~~~~~~else (nR\_AND X L3)) (\{0; 1\; 2\}) t =}}}\\
\mbox{\textup{\texttt{~~p\_space p DIFF}}}\\
\mbox{\textup{\texttt{~~DFT\_event p }}}\\
\mbox{\textup{\texttt{~~~(n\_OR}}}\\
\mbox{\textup{\texttt{~~~~(MAP}}}\\
\mbox{\textup{\texttt{~~~~~~($\lambda$i.}}}\\
\mbox{\textup{\texttt{~~~~~~~~~~~if i = 0 then WSP Y Ys\textsubscript{a} Ys\textsubscript{d}}}}\\
\mbox{\textup{\texttt{~~~~~~~~~~~else if i = 1 then}}}\\
\mbox{\textup{\texttt{~~~~~~~~~~~~~~~~\big((n\_OR (MAP X (SET\_TO\_LIST L1))) $\cdot$}}}\\
\mbox{\textup{\texttt{~~~~~~~~~~~~~~~~~(n\_OR (MAP X (SET\_TO\_LIST L2)))\big)}}}\\
\mbox{\textup{\texttt{~~~~~~~~~~~else (n\_OR (MAP X (SET\_TO\_LIST L3))))}}}\\
\mbox{\textup{\texttt{~~~~~~~~~~(\{0; 1\; 2\}))) t)}}}
\end{theorem}

It is worth mentioning that Theorem~\ref{thm:sen_eq} can be used for the equivalence of the DRBD-DFT models of the SEN in both the terminal and broadcast since they both share the same structure.

Based on these theorems, we can use one model to verify the probability of the other model using the probability of the complement.

\section{Conclusion}
In this report, we presented the formal dynamic dependability analysis of SEN and SEN+ MINs that form a critical part in the routing process of multiprocessor systems. We provided generic expressions of reliability and probability of failure that are independent of the failure distributions. Furthermore, we verified these expressions for an arbitrary number of system blocks that can be instantiated later to a certain number without the need to repeat the verification process. For instance, we evaluated the reliability and probability of failure using MATLAB for a specific number of system components based on these generic expressions. It is worth mentioning that such sound generic results cannot be obtained using simulation or model checking as the state space should be defined in advance. The proof script of the verification of SEN and SEN+ is available at~\cite{SEN-code} and it took around 80 hours to be developed.


\bibliographystyle{unsrt}

\bibliography{mabiblio}
\end{document}